%% file: thesis.tex
\documentclass[logo,bsc,singlespacing,parskip,online]{infthesis}

\usepackage{ugcheck}
\usepackage{microtype} 
\usepackage[numbers]{natbib} 
\usepackage{graphicx}
\usepackage{float}
\usepackage{booktabs}
\usepackage{amsmath}
\usepackage{amssymb}
\usepackage{bbm}
\usepackage{url}
\usepackage{hyperref}
\usepackage{cleveref}
\begin{document}

\input{chapters/preliminaries}
\input{chapters/introduction}
\input{chapters/background}

\input{chapters/experimental_setup}
\input{chapters/crnn}
\input{chapters/model_improvements}
\input{chapters/conclusions}

\bibliographystyle{unsrtnat}
\bibliography{references}

\input{chapters/appendix}

\end{document}

%% file: chapters/preliminaries.tex
\begin{preliminary}

    \title{Chord Recognition with Deep Learning}
    
    \author{Pierre Lardet}
    
    \course{Computer Science and Mathematics}
    
    \project{4th Year Project Report}

    \date{\today}
    
    \abstract{
      Progress in automatic chord recognition has been slow since the advent of deep learning in the field. To understand why, I conduct experiments on existing methods and test hypotheses enabled by recent developments in generative models. Findings show that chord classifiers perform poorly on rare chords and that pitch augmentation boosts accuracy. Features extracted from generative models do not help and synthetic data presents an exciting avenue for future work. I conclude by improving the interpretability of model outputs with beat detection, reporting some of the best results in the field and providing qualitative analysis. Much work remains to solve automatic chord recognition, but I hope this thesis will chart a path for others to try.
    }
    
    \maketitle
    
    \newenvironment{ethics}
       {\begin{frontenv}{Research Ethics Approval}{\LARGE}}
       {\end{frontenv}\newpage}
    
    \begin{ethics}
    This project was planned in accordance with the Informatics Research
    Ethics policy. It did not involve any aspects that required approval
    from the Informatics Research Ethics committee.
    
    \standarddeclaration
    \end{ethics}

    \begin{acknowledgements}
    Thank you to my two lovely supervisors, \hyperlink{https://homepages.inf.ed.ac.uk/amos/index.html}{Prof.~Amos Storkey} and \hyperlink{https://www.acoustics.ed.ac.uk/people/dr-alec-wright/}{Dr.~Alec Wright} for allowing me to fulfil my desire for such a personally motivating project. They offered guidance on this project and research more broadly, which I will take forward. I would like to thank Andrea Poltronieri for sharing the dataset used in this work. Finally, I would like to thank my friends and family for giving me reasons to take a break.
    \end{acknowledgements}

    \tableofcontents
    \end{preliminary}

%% file: chapters/introduction.tex
\chapter{Introduction}

Chords form an integral part of music. Part of how musicians understand music is through harmonic structure. Chord annotations are a symbolic representation of the chords in a piece of music. They allow music to be easily shared, performed, improvised and analysed. Not all chord annotations available online are free or of a high enough quality because creating high-quality chord annotations requires a trained musician. 

To this end, I investigate the use of deep learning in automatic chord recognition to create chord annotations of music. Data-driven methods have dominated the field for over a decade. However, the significant progress of early models has not continued in recent years; the problem remains far from solved.

In this work, I first aim to understand why performance improvements have stagnated. I implement a standard benchmark model and conduct a thorough analysis of its behaviour. This involves looking at the model's common mistakes, performance on rarer chords and how its predictions relate to time. I then use these observations to study different methods of improving these models. 

I also conduct novel research on using generative models as both feature extractors and a source of new data. This is enabled by chord-conditioned generative models developed in recent years. I conclude by rethinking how the model predicts chords in time by incorporating beat estimation. 

This work goes towards enabling software which can be used to better understand, create and learn music. Easily accessible and accurate chord recognition models would allow producers to better understand their work and musicologists to study larger datasets. Musicians and hobbyists could access chord annotations for their favourite songs or analyse their performances and improvisations.

The analysis of existing models, exploration of improvements and discussion of new research directions constitute a novel contribution to the field of automatic chord recognition. Despite the lack of performance improvements in recent years, I hope this work motivates others to continue pursuing research aiming to solve the problem posed by automatic chord recognition. 

\section{Outline}

The thesis is structured as follows:

\begin{itemize}
    \item \textbf{Chapter 2} provides background information on harmony, chord recognition and musical data. I then discuss existing literature on the subject, pointing out trends in the field and the most exciting avenues for research.
    \item \textbf{Chapter 3} describes the datasets, evaluation metrics and training procedure used.
    \item \textbf{Chapter 4} contains the implementation of a convolutional neural network from the literature, followed by an analysis of its properties and predictions. I observe behaviours which provide opportunities to improve the model. 
    \item \textbf{Chapter 5} extends this work by studying various methods of improvement. Some of these experiments analyse existing improvements, while others present novel avenues of research.
    \item \textbf{Chapter 6} concludes the thesis and provides suggestions for future work.
\end{itemize}

\vspace{0.5cm}

All code is available on GitHub.\footnote{\url{https://github.com/PierreRL/AutomaticChordRecognition}} Data can be made available upon request.\footnote{lardet[dot]pierre[at]gmail.com}

%% file: chapters/background.tex
\chapter{Background \& Related Work}

In this chapter, I first introduce harmony and chords and their role in music. I then discuss how music can be represented as input to a machine learning model. This is followed by an overview of the field of automatic chord recognition (ACR). This includes the datasets and models that are commonly used in ACR, the challenges that are faced in this field and future directions.

\section{Background}

\subsection{Harmony and Chords}

Harmony is the combination of simultaneously sounded notes. A common interpretation of such sounds is as a chord. Chords can be thought of as a collection of at least two notes built from a root note and scale. Any notes from the scale can be present, but the most common are the third, fifth and seventh. A chord's \emph{quality} is determined by the intervals between notes in the chord irrespective of the root note. The most common qualities are major and minor. Many other qualities exist, such as diminished and suspended. Chords can be played in \emph{inversion}, where the root note is not the lowest. In this work, chords are represented using Harte notation~\citep{HarteNotation}.

Chords can be closely related. \texttt{C:maj7} is very close to \texttt{C:maj}. The only difference is an added major seventh. An important relation in music theory is between \emph{relative major/minor} chords. These pairs of chords are built from the same scale and share many notes. For example, \texttt{G:maj} and \texttt{E:min} are related in this way. It is possible for different chords to share the same set of pitch classes like \texttt{G:maj6} and \texttt{E:min7}.

Chords are an important part of music. They provide harmonic context for a melody and can be used to convey emotion, tension and release~\citep{HarmonyandVoiceLeading}. They are also crucial for improvisation, where musicians will play notes that fit the chord progression~\citep{JazzTheoryBook}. Contemporary guitar music is often represented by a chord sequence. Chords are also important for songwriting and production, where a chord progression can form the basis of a song. Music analysis also makes heavy use of chords. Musicologists can analyse the harmonic structure of a piece to understand the composer's intentions and why we enjoy the music~\citep{TonalAnalysisBach}.

\subsection{Chord Recognition}

Chord recognition is the task of identifying which chord is playing at any moment in a piece of music. This can be useful for creating notated versions of songs for musicians, musicologists and music recommendation. Those wishing to learn a song may visit websites such as Ultimate Guitar\footnote{\url{https://www.ultimate-guitar.com/}} where users submit chord annotations for songs. Musicologists may wish to analyse the harmonic structure of a piece of music. Music recommendation systems can recommend songs based on their harmonic content as similar music will often have similar harmonic content~\citep{MusicGenreClassification}. For example, modern pop music famously uses many similar chords~\footnote{\url{https://www.youtube.com/watch?v=oOlDewpCfZQ} accessed 25th February 2025} while contemporary jazz music is known for its complex and rich exploration of harmony.

All of the above motivate the need for accurate chord annotations. However, annotations from online sources can be of varying quality and may not be available for all songs~\citep{Choco}. The task of annotating chords is time-consuming and requires a trained musician~\citep{McgillBillboard}. Automatic chord recognition systems have the potential to alleviate these problems by providing a fast, accurate and scalable solution.

Unfortunately, chord recognition is a non-trivial task. Which chord is playing is inherently ambiguous. Different chords can share the same notes. The same chord can be played on different instruments with unique timbres. Precisely when a chord starts and ends can be imprecise. Whether a melody note is part of a chord and whether a melody alone is enough to imply harmonic content are both ambiguous. In order to identify a chord, data across time must be considered. For example, a chord may be vamped or arpeggiated. Audio also contains many unhelpful elements for chord recognition such as reverb, distortion and unpitched percussion.

\subsection{Music Features}\label{sec:background-features}

Recorded music can be represented in a variety of ways on a computer. The simplest is as a raw time series of amplitudes, referred to as the audio's waveform. Data in the raw audio domain has been applied in generative models such as Jukebox~\citep{Jukebox} and MusicGen~\citep{MusicGen} and autoencoders~\citep{Encodec}.

\textbf{Spectrogram}: A spectrogram is a transformation of a waveform into the time-frequency domain calculated via a short-time Fourier transform (SFTF). Spectrograms are commonly used in many audio processing tasks such as audio search~\citep{ShazamSpectrogram} and music transcription~\citep{PianoTranscriptionWithTransformer}. As of yet, linear spectrograms computed using the STFT have not been used in ACR tasks~\citep{20YearsofACR}. Other transformations relate better to how humans understand pitch and harmony.

\begin{figure}[h]
    \centering
    \includegraphics[width=0.8\textwidth]{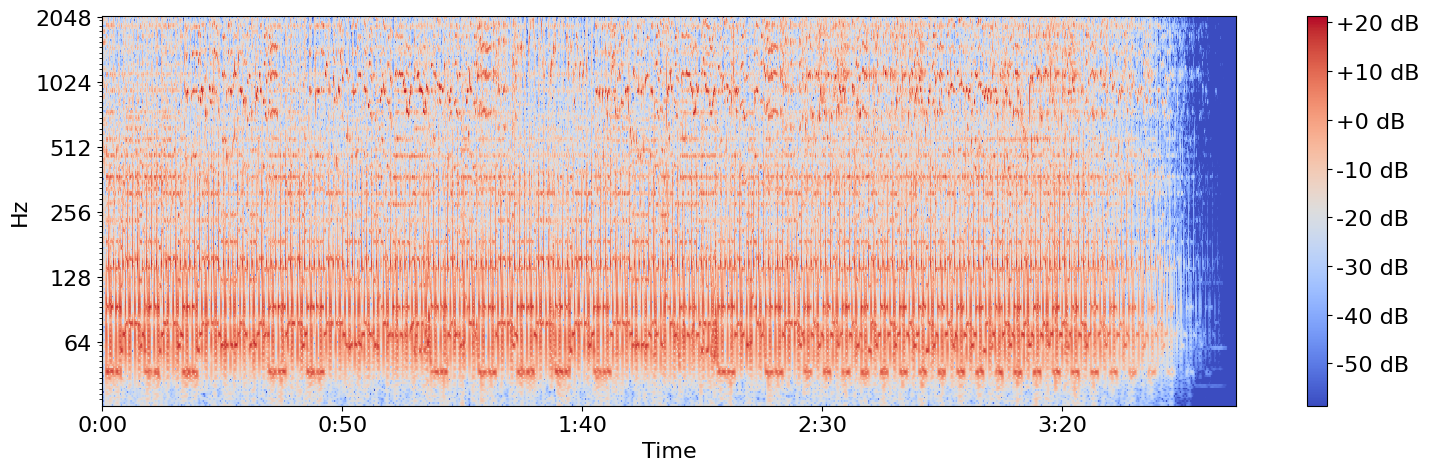}
    \caption{CQT of `Girls Just Wanna Have Fun' by Cyndi Lauper. We can see the log-spaced frequency bins on the y-axis. There is clear structure and repetition in the song, particularly in the lower frequencies, which can be attributed to a regular drum groove. Such structure gives an idea of the patterns a machine learning model may look for to identify chords.}~\label{fig:cqt_example}
\end{figure}

\textbf{CQT}: A common version of the spectrogram used in music transcription is the constant-Q transform (CQT)~\citep{CQT}. The frequency bins of a CQT are logarithmically spaced and has bin widths that are proportional to frequency. This is motivated by the logarithmic nature of how humans perceive pitch intervals in music: a sine wave with double the frequency is perceived as one octave higher. As such, CQTs are used in many music transcription tasks and are very popular for ACR~\citep{FirstDeepLearningCQT,StructuredTraining}. The \emph{hop length} of the CQT is the number of samples used from the waveform to calculate representations per frame being analysed. It determines the time resolution of the CQT. A shorter hop length results in a higher time resolution. An example CQT from the dataset used in this work is shown in Figure~\ref{fig:cqt_example}. As \citet{SaliencyChroma} note, CQTs are preferred to other spectrograms for ACR due to their finer resolution at lower frequencies and for the ease with which pitch can be studied and manipulated. For example, CQTs make pitch shifting possible through a simple shift of the CQT bins. Another logarithmic variant of the spectrogram is the mel-spectrogram, based on the mel-scale~\citep{MelScale}. It is intended to mimic the human ear's perception of sound and is commonly used in speech recognition~\citep{SpeechProcessingMels} but has also been used in music transcription tasks~\citep{MelodyTranscriptionViaGenerativePreTraining}.


\textbf{Chroma Vectors}: Chroma vectors are a 12-dimensional time-series representation where each dimension corresponds to a pitch class. Each element represents the strength of each pitch class in the Western chromatic scale. Such features have been generated by deep learning methods~\citep{BalanceRandomForestACR} or by hand-crafted methods~\citep{NNLSChroma,librosa} and have seen use in recent ACR models~\citep{HarmonyTransformer}. A representation of a song as a chroma vector over time can be thought of as another type of spectrogram, referred to as a \emph{chromagram}.

\textbf{Generative Features}: More recently, features extracted from generative music models have been used as input, referred to in this work as \emph{generative features}. The proposed benefit is that the vast quantities of data used to train these models leads to rich representations of the music. These features have been shown to contain useful information for music information retrieval (MIR) tasks~\citep{GenerativeFeaturesforMIR}. \citet{MelodyTranscriptionViaGenerativePreTraining} use features from JukeBox~\citep{Jukebox} to train a transformer~\citep{AttentionIsAllYouNeed} for both melody transcription and chord recognition. They found that these features outperformed mel-spectrograms in melody transcription tasks but did not report results for ACR nor with CQTs.

\section{Related Work}\label{sec:related_work}

The field of ACR has seen considerable research since the seminal work of \citet{FujishimaACR} in 1999. Below, I provide a brief overview of the field including the datasets, metrics and models that are commonly used. I conclude by discussing some of the common challenges faced and motivating the research carried out in this project. 

\subsection{Data}\label{sec:background-data}

Sources of data that have seen common use in ACR relevant to this work include:

\begin{itemize}
    \item \emph{Mcgill Billboard}: over 1000 chord annotations of songs randomly selected from the Billboard `Hot 100' Chart between 1958 and 1991.~\citep{McgillBillboard}
    \item \emph{Isophonics}: 300 annotations of songs from albums by The Beatles, Carole King and Zweieck.~\citep{Isophonics}
    \item \emph{RWC-Pop}: 100 pop songs with annotations available\footnote{\url{https://github.com/tmc323/Chord-Annotations}} for chords.~\citep{RWC}
    \item \emph{USPop}: 195 annotations of songs chosen for popularity.~\citep{USPop}
    \item \emph{JAAH}: 113 annotations of a collection of jazz recordings.~\citep{JAAH}
    \item \emph{HookTheory}: 50 hours of labelled audio in the form of short musical segments, crowdsourced from an online forum called HookTheory\footnote{https://www.hooktheory.com/}.~\citep{MelodyTranscriptionViaGenerativePreTraining}
\end{itemize}

Many of these have been compiled into the \emph{Chord Corpus} by \citet{Choco} with standardised annotation formats. However, audio is scarce due to copyright issues. The most common dataset is comprised of 1217 songs compiled from the first four of the above collections. This dataset is dominated by pop songs.

Another problem is that the existing data is imbalanced, with a large number of major and minor chords and fewer instances of chords with more obscure qualities. This can lead to models that are biased towards predicting major and minor chords. Attempts to address such an imbalanced distribution have been made by weighting the loss function~\citep{ACRLargeVocab1}, adding `structure' to the chord targets~\citep{StructuredTraining,ACRLargeVocab1}, re-sampling training examples to balance chord classes~\citep{BalanceRandomForestACR} and curriculum learning~\citep{CurriculumLearning}.

\textbf{Pitch Augmentation}: Due to the lack of labelled data, data augmentation via pitch shifting has been applied to ACR. Input features are pitch shifted while chords are transposed. \citet{StructuredTraining} note the large increase in performance using pitch shifting directly on the audio. Other works have since used pitch shifting directly on the CQT~\citep{ACRLargeVocab1}. No work has compared the two methods.

\textbf{Synthetic Data Generation}: Data has been scaled up using augmentation and semi-supervised learning with some success~\citep{ScalingUpSemiSupervisedLearning}. Research has been done into the use of synthetic data~\citep{MusicGenTrainingData,AnnotationFreeSyntheticData} and self-supervised learning~\citep{MERTSupervisedLearning} for other MIR tasks but not for ACR. With the advent of new models which accept chord-conditioned input~\citep{MusiConGen}, the possibility of generating synthetic data for ACR is an exciting avenue of research.





\subsection{Models}

\textbf{Model Architectures}: Since the work of \citet{RethinkingChordRecognition}, chord recognition has been predominantly tackled by deep learning architectures. The authors used a convolutional neural network (CNN) to classify chords from a CQT. CNNs have been combined with recurrent neural networks (RNNs)~\citep{ACRCNNRNN1,ACRLargeVocab1,StructuredTraining} with a CNN performing feature extraction from a spectrogram and an RNN sharing information across frames. More recently, transformers have been applied in place of the RNN~\citep{MelodyTranscriptionViaGenerativePreTraining, HarmonyTransformer, AttendToChords,CurriculumLearning,BTC}.

Despite increasingly complex models being proposed, performance has not improved by much. In fact, \citet{BTC} found that their transformer performed marginally worse than a CNN. \citet{FourTimelyInsights} talk of a 'glass ceiling' with increases in performance stagnating after the advent of deep learning in ACR. This was 10 years ago and the situation has not changed significantly. Despite this, continued efforts have been made to develop complex models with the sole motivation of improving performance. This has led to overly complex ACR models seeing use in other MIR tasks such as chord-conditioned generation where \citet{MusiConGen} use the model developed by \citet{BTC} despite its lack of improvement over simpler predecessors. Furthermore, there is little comparison to simple baselines to provide context for the performance gain associated with increasing model complexity.

\textbf{Decoding}: A decoding step is often performed on the probabilities outputted by the neural network. This can smooth predictions and share information across frames. \citet{BalanceRandomForestACR} use a hidden Markov model (HMM), treating the probability distributions over chords generated by the model as emission probabilities and constructing a hand-crafted transition function. Other works have used conditional random fields (CRF) to model the dependencies between chords~\citep{ACRLargeVocab1}. Both HMMs and CRFs can use either learned transition matrices or homogeneous penalties for transitions to different chords. It is unclear whether or not learning transitions is better. In both cases, self-transition probabilities are very large and \citet{CommonVariations} argue that increases in performance can be mostly attributed to the reduction in the number of transitions. However, a more recent analysis of such behaviour is missing from the literature.

\textbf{Model Analysis}: \citet{FeatureMaps} visualise the outputs of layers of the CNN and find that some feature maps correspond to the presence of specific pitches and intervals. \citet{SaliencyChroma} visualise the importance of different parts of an input CQT using saliency maps, noting the clear correlation between pitch classes present in a chord and the saliency maps. Confusion matrices over chord roots and qualities are also commonly used to analyse the performance of models. For example, \citet{StructuredTraining} found that similar qualities are often confused with each other and that the model favours the most common chord qualities. ~\citet{BTC} attempt to interpret attention maps produced by their transformer as musically meaningful.

Regardless of such analyses, too much effort is spent on motivating complex model architectures with a focus on minor improvements in performance. In this work, I will conduct a thorough analysis of an existing model. I will take inspiration from some of the analyses above while adding a more nuanced understanding of the model's behaviour and failure modes by way of example.

\subsection{Frames and Beats}

Chords exist in time. How the time dimension is processed prior to being fed into the model matters. When audio is transformed into a spectrogram, each vector of frequencies represents a fixed length of time, called a \emph{frame}. The frame length is determined by the hop length used when calculating the CQT. Constant frame lengths can be made short enough such that the constraint imposed on the model to output chord predictions on a per-frame basis is not limiting. However, different hop lengths have been used, varying from 512~\cite{ACRLargeVocab1} up to 4096~\citep{StructuredTraining}. Which hop length works best remains unclear.

More recently, \citet{MelodyTranscriptionViaGenerativePreTraining} used a frame length determined by beats detected from the audio. Because they focus primarily on melody transcription, they define frames to be a 1/16th note $\approx 125$ms with 120 beats per minute (BPM). Such beat synchronicity has been proposed for chord recognition. The underlying assumption is that chords tend to change on the beat. This reduces the computational cost of running the model due to a decreased frame rate and, more importantly, leads to a more musically meaningful interpretation of the output. However, \citet{CommonVariations} and \citet{RelativePerformance} argue that because beat detection is far from perfect, restricting frames to beats can hurt performance. Beat detection models have improved since then. Proper analysis of beat-synchronous chord recognition in the modern setting is lacking in the literature.  \citet{ChorusAlignmentJAAH} jointly estimate beats and chords but use a different jazz-specific dataset and do not analyse how beat-wise predictions affect performance.

\subsection{Future Directions}

\citet{20YearsofACR} provide an overview of ACR up to 2019 since the seminal work of \citet{FujishimaACR} in 1999 and provide suggestions for future avenues of research. They look at future research directions. This includes the use of different representations for both audio and chords, addressing the mismatch between chord changes and discretised frames fed to a model, looking at the larger structures in music like verses and chords, incorporating other elements of the music such as melody and genre, methods of handling subjectivity of chords and the imbalance present in chord datasets. Since then, different works have addressed some of these problems in various ways. Among these problems, the focus has been primarily on addressing the imbalance in the chord dataset. 

In this work, I will implement a simple model that remains competitive with the state-of-the-art~\citep{StructuredTraining}. I will then conduct a thorough analysis of the model and its architecture. I will look at common methods for improving ACR models with more detailed analyses than have previously been conducted. This analysis will provide insight into the strengths and weaknesses of such models. It may also provide guidance for further improvements. I will also look at novel methods of improvement made possible through generative and beat detection models. This includes the use of generative features and synthetic data as input to the model as well as beat-synchronous frames. Finally, I will evaluate the improved models in terms of their performance and as a tool for musicians and musicologists.

%% file: chapters/experimental_setup.tex
\chapter{Experimental Setup}

In this chapter, I outline the datasets used in this work, the preprocessing applied to the audio and chord annotations, the evaluation metrics used to compare the models and details of the training process.

\section{Data}

I spent the initial period of this project finding a suitable dataset for training and testing. \citet{JAAH} use the \emph{JAAH} dataset while \citet{MelodyTranscriptionViaGenerativePreTraining} use the \emph{HookTheory} dataset. Many works use a combination of the \emph{McGill Billboard}, \emph{Isophonics}, \emph{RWC-Pop} and \emph{USPop} datasets. However, none have audio available. Furthermore, annotations come from different sources in different formats. I spent time looking at other sources audio data, pre-computed features of audio which are available for some datasets and compiling disparate annotations in different formats. I also contacted authors of previous ACR works to see if they could provide me with audio. I was able to get in contact with Andrea Poltronieri, a PhD student at the University of Bologna, one of the authors of the chord corpus or 'ChoCo' for short~\citep{Choco}. He provided me with labelled audio for the 1217 songs that are commonly used, alongside labelled audio for the \emph{JAAH} dataset. This was a great help despite it coming several weeks into the project.

I refer to the dataset used in this work as \emph{pop}. It consists of songs from the \emph{Mcgill Billboard}, \emph{Isophonics}, \emph{RWC-Pop} and \emph{USPop} datasets introduced in Section~\ref{sec:background-data}. This collection was initially proposed in work by \citet{FourTimelyInsights} to combine some of the known datasets for chord recognition. The dataset consists of subsets from the above sources filtered for duplicates and selected for those with annotations available. In total, there are 1,213 songs. The dataset was provided with obfuscated filenames and audio as \texttt{mp3} files and annotations as \texttt{jams} files~\citep{JAMS}. 

\newpage
\subsection{Data Integrity}\label{sec:data-integrity}

Several possible sources of error in the dataset are investigated below.

\textbf{Duplicates:} Files are renamed using provided metadata identifying them by artist and song title. This is done to identify duplicates in the dataset. There is only one: Blondie's \emph{One Way or Another}, which has two different recordings. It is removed from the dataset. Automatic duplicate detection is conducted by embedding each audio using mel-frequency cepstral coefficients (MFCC)~\citep{MFCC}. This function is commonly used to embed audio into low dimensions. This provides a fast and easy way of quantifying similarity. Audio is passed through the \texttt{mfcc} provided in \texttt{librosa}~\citep{librosa} with 20 coefficients. A song's embedding is calculated as the mean MFCC over all frames. Cosine similarities are then calculated for all pairs of tracks. None of the top 50 similarity scores yielded any sign of duplication. I proceed with the assumption that there are no further duplicates in the dataset.

\textbf{Chord-Audio Alignment:} It is pertinent to verify that the chord annotations align with the audio. Misaligned annotations could make training impossible. Ten songs are manually investigated for alignment issues by listening to the audio and comparing it to the annotations directly. The annotations are all well-timed with detailed chord labels.

Automatic analysis of the alignment of the audio and chord annotations is also done using the cross-correlation between the derivative of the CQT over time and the chord annotation. A maximum correlation at a lag of zero would indicate good alignment as the audio changes at the same time as the annotation. The derivative of the CQT in the time dimension is estimated using the \texttt{delta} function from \texttt{librosa}. The chord annotations are converted to a binary vector, where each element corresponds to a frame in the CQT and is 1 if a chord change occurs at that frame and 0 otherwise. Both the CQT derivatives and binary vectors are normalised by subtracting the mean and dividing by the standard deviation. Finally, cross-correlation is computed using the \texttt{correlate} function from \texttt{numpy}. A typical cross-correlation for a song is visualised in Appendix~\ref{app:cross_correlation}. The cross-correlation is periodic and repeats every 20 frames or so. Listening to the song, the period of repetition is a fraction of a bar length. 

\begin{figure}[H]
    \centering
    \includegraphics[width=1.0\textwidth]{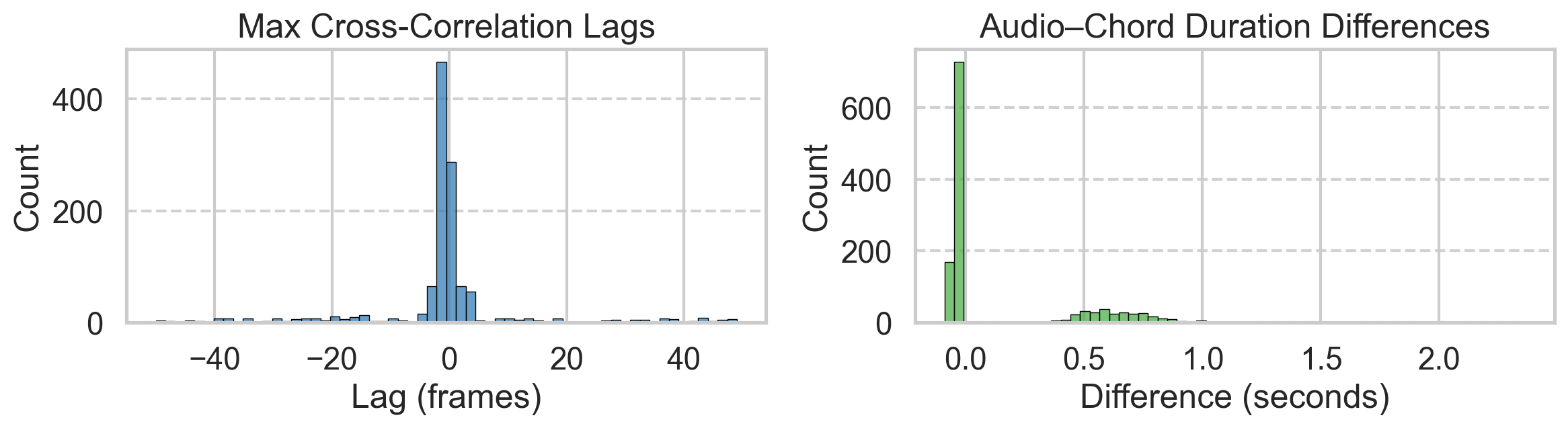}
    \caption{Histograms of the maximum cross-correlation lags and the difference in length between the audio and chord annotations. Both show results close to $0$, suggesting good alignment between audio and annotations. }\label{fig:durations-and-lags}
\end{figure}

To check alignment across the dataset, I plot a histogram of the lag of the maximum cross-correlations over songs in Figure~\ref{fig:durations-and-lags}. Under the assumption that the annotations are not incorrect by more than 5 seconds, the region of possible maximal lags is restricted to a window of 50 frames on either side of 0. This reduction does not change the shape of the picture. Instead, focusing on a reduced set of lags allows more detail to be visible. The majority of songs have a maximum lag close to 0, with a few outliers. This can be attributed to noise. A final check is done by looking at the difference in length of the audio files and chord annotations. A histogram of differences in length is also shown in the figure. The majority of songs have a difference in length of 0, with a few outliers almost all less than a second. This evidence, combined with the qualitative analysis, is convincing enough to leave the annotations as they are for training.

\textbf{Incorrect and Subjective Annotations:} Throughout manual listening, no obviously wrong annotations were found. However, looking at songs on which the preliminary models perform the worst using the \texttt{mirex} metric, three songs stick out. `Lovely Rita' by the Beatles, `Let Me Get to Know You' by Paul Anka and `Nowhere to Run' by Martha Reeves and the Vandellas all had scores below $0.05$. In these songs, the model consistently guessed chords one semitone off, as if it thought the song was in a different key. Upon listening, it became clear that the tuning was not in standard A440Hz for the first two songs and the key of the annotation was wrong for the other. These songs were removed from the dataset. All reported results exclude these data points. No other songs were found to have such issues.

Chord annotations are inherently subjective to some extent. Detailed examples in \emph{pop} are given by \citet{FourTimelyInsights}. They also note the presence of several songs in the dataset of questionable relevance to ACR, as the music itself is not well-explained by chord annotations. However, these are kept in for consistency with other works as this dataset is often used in the literature. Some works decide to use the median as opposed to the mean accuracy in their evaluations in order to counteract the effect of such songs on performance~\citep{StructuredTraining}. We think that this is unnecessary as the effect of these songs is likely to be small and we do not wish to inflate our results inadvertently. Further evidence for the use of the mean is given in Section~\ref{sec:evaluation}.

\subsection{Preprocessing}

\subsubsection{Audio to CQT}\label{sec:audio-to-cqt}

I first convert the audio to a constant-Q transform (CQT) representation introduced in Section~\ref{sec:background-features}. CQTs are common in ACR and is used as a starting point for this work. The CQT was computed using \texttt{librosa}, using the built-in \texttt{cqt} function. A sampling rate of $44100$Hz was used, with a hop size of 4096, 36 bins per octave, 6 octaves and a fundamental frequency corresponding to the note \texttt{C1}. These parameters were chosen to be consistent with previous works~\citep{StructuredTraining} and with standard distribution formats. The CQT is returned as a complex-valued matrix containing phase, frequency and amplitude information. Phase information was discarded by taking the absolute value before being converted from amplitude to decibels (dB), equivalent to taking the logarithm.

The CQT matrix of a song has size $216 \times F$ where 216 is the number of frequency bins and $F$ is the number of frames in the song. The number of frames can be calculated as $F = \lceil \frac{44100}{4096} L  \rceil$ where $L$ is the length of the song in seconds, 44100 is the sampling rate in Hertz (Hz) and 4096 is the hop length in samples. A 3-minute song has just under 2000 frames. To save on computational cost, the CQT was pre-computed into a cached dataset rather than re-computing each CQT on every run.

\subsubsection{Chord Annotations}\label{sec:chord-annotations}

The chord annotation of a song is represented as a sorted dictionary, where each entry contains the chord label, the start time and the duration. The chord label is represented as a string in Harte notation~\citep{HarteNotation}. For example, C major 7 is \texttt{C:maj7} and A half diminished 7th in its second inversion is \texttt{A:hdim7/5}. The notation also includes \texttt{N}, a no chord symbol, and \texttt{X}, an unknown chord symbol.

Such a chord vocabulary is too flexible to be used as directly as a target for a machine learning classifier trained on limited data. It would contain thousands of classes, many of which would appear only once. Instead, I define a restricted chord vocabulary. This contains 14 qualities: major, minor, diminished, augmented, minor 6, major 6, minor 7, minor-major 7, major 7, dominant 7, diminished 7, half-diminished 7, suspended 2, and suspended 4. \texttt{N} denotes no chord playing and chords outside the vocabulary are mapped to \texttt{X}, a dedicated unknown symbol. Letting $C$ denote the size of the chord vocabulary, $C=12\cdot14 + 2 = 170$. This vocabulary is consistent with much of the literature~\citep{StructuredTraining,FourTimelyInsights,ACRLargeVocab1}. \citet{ACRLargeVocab1} use a more detailed vocabulary by including inversions, but I decide to remain consistent with other previous works. As \citet{StructuredTraining} note, $C=170$ is sufficient for the dataset to exhibit a significant imbalance in the chord distribution. Their methodology is easily extensible to larger vocabularies. If performance is not yet satisfactory on $C=170$, it is unlikely that performance will improve with a larger vocabulary.

Both training labels and evaluation labels are converted to this vocabulary. If the evaluation labels were kept in the original Harte notation, the model would be unable to identify them. The method for converting from Harte notation to a symbol in the chord vocabulary is similar to that used by \citet{StructuredTraining} and is detailed in Appendix~\ref{app:chord_mapping}.

A simpler chord vocabulary is also sometimes used. It contains only the major and minor quality for each root, and the \texttt{N} and \texttt{X} symbols. For example, \texttt{C:maj7} is mapped to \texttt{C:maj} while \texttt{A:hdim7/5} is mapped to \texttt{X}. For this vocabulary, $C=26$. I did some preliminary tests with this vocabulary but quickly found that model performance was similar over the two vocabularies. Results and analysis can be found in Appendix~\ref{app:small_vs_large_vocabulary}. Additionally, the \texttt{majmin} evaluation metric compares chords over this smaller vocabulary and is mentioned in Section~\ref{sec:evaluation}. The smaller vocabulary is not used in the rest of this work as there seems to be no advantage over the larger vocabulary.

Frames are allocated a chord symbol based on which chord is playing in the middle of the frame. While this may not be a perfect solution, frames are $\approx93$ms long, which is shorter than the minimum duration of a chord in the dataset. This guarantees that the chord label for every frame plays for at least half the frame. Furthermore, only $4.4\%$ of frames include a chord transition.

\subsection{Chord Distribution}

\begin{figure}[H]
    \centering
    \includegraphics[width=1.0\textwidth]{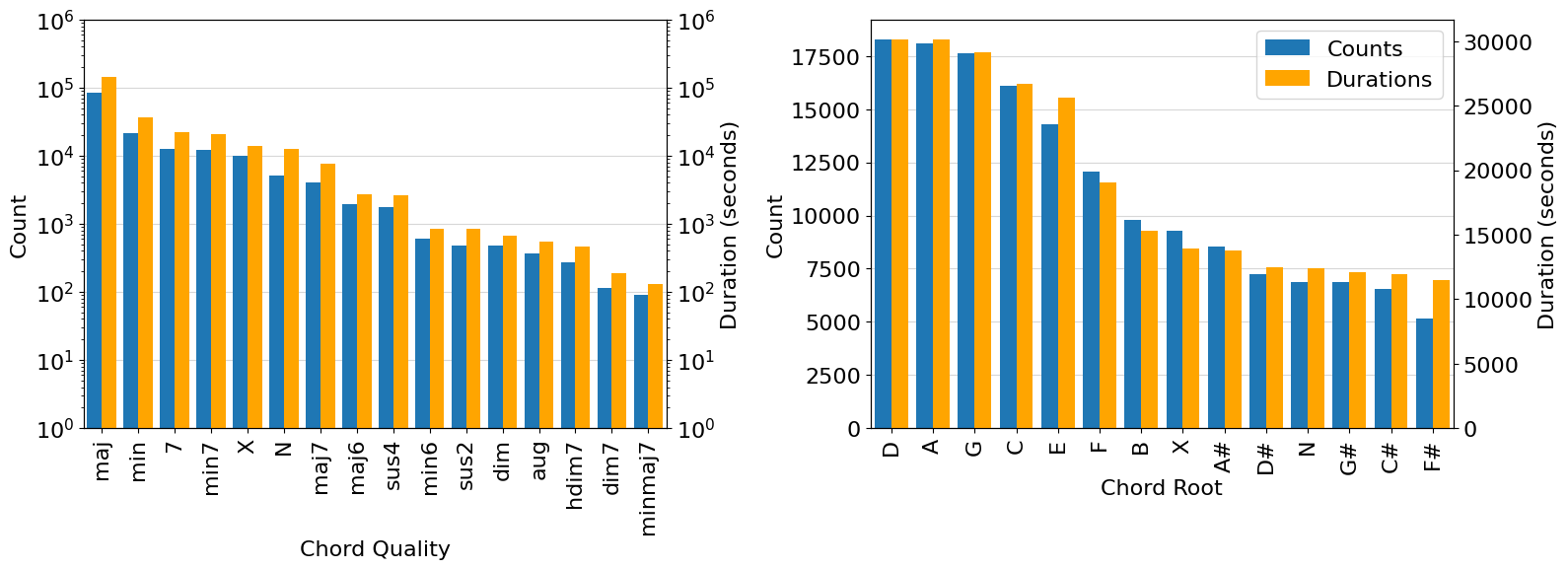}
    \caption{Chord distributions over qualities (left) and roots (right) in the \emph{pop} dataset. The plots show the raw counts in frames and the duration in seconds for each chord root/quality. Note that the y-axis over qualities is in a logarithmic scale. The qualities are very imbalanced, with \texttt{maj} as the most popular. Conversely, roots are relatively balanced.}\label{fig:chord-distribution}
\end{figure}

Much of the recent literature has focused on the long tail of the chord distribution, using various methods to address the issue. It is first helpful to understand the distribution of chords in the datasets, illustrated in Figure~\ref{fig:chord-distribution}. The distribution is broken down both by root and quality, using the chord vocabulary with $C=170$. The plots show that the distribution over qualities is highly skewed, with major and minor chords making up the majority of the dataset, and qualities like majorminor and diminished 7th chords, two or three orders of magnitude less common. The distribution over roots is far less skewed, although there is a preference for chords in keys with common roots like A, C and D.

\section{Evaluation}\label{sec:evaluation}

As is standard for ACR, I use weighted chord symbol recall (WCSR) to evaluate chord classifiers. Simply put, WCSR measures the fraction of time that a classifier's predictions are correct. I include a formal definition in Appendix~\ref{app:weighted_chord_symbol_recall_definitions}. Correctness can be measured in a variety of ways, such as \texttt{root}, \texttt{third} and \texttt{seventh}, which compare along roots, thirds, or sevenths respectively. I also use the \texttt{mirex} score, where a prediction is correct if it shares at least three notes with the label. This allows for errors like mistaking \texttt{C:7} for \texttt{C:maj}. Finally, I use \texttt{acc}, or simply accuracy, to denote the overall accuracy where symbols must match exactly.

Other measures of correctness are sometimes used. These include \texttt{majmin}, a measure of correctness over only major and minor qualities. I use this measure only to substantiate the use of the larger vocabulary in Appendix~\ref{app:small_vs_large_vocabulary}. Measures of correctness over triads and tetrads are also sometimes used, but these are highly correlated with \texttt{third} and \texttt{seventh}, respectively. This correlation is to be expected as the third and seventh are strong indicators of the triad and tetrad of the chord. This was verified empirically on preliminary experiments which are omitted.

All metrics are implemented in the \texttt{mir\_eval} library~\citep{mir_eval}, which also provides utilities for calculating WCSR from frame-wise chord predictions. The mean WCSR is computed over all songs in the evaluation set. Some other works report the median. Empirically, I found the median to be $\approx 2\%$ greater than the mean. This may be due to those songs identified as being unsuitable for chordal analysis by \citet{FourTimelyInsights}. I report only the mean throughout this work. This was chosen for being more commonly used in recent literature and because it is important for a metric to detect if the model performs poorly over certain genres.

For some experiments, two further metrics are calculated. These are the mean and median class-wise accuracies, called \texttt{acc}\textsubscript{class} and \texttt{median}\textsubscript{class}, respectively. \texttt{acc}\textsubscript{class} has previously been defined in terms of discrete frames by \citet{ACRLargeVocab1}. I redefine \texttt{acc}\textsubscript{class} here in terms of WCSR and introduce \texttt{median}\textsubscript{class}. The definitions can be found in Equations~\ref{eq:class_wise}.

\begin{equation}\label{eq:class_wise}
    \text{acc}\textsubscript{class} = \frac{1}{C} \sum_{c=1}^{C} \text{\emph{WCSR}}(c)
\quad \quad
    \text{median}\textsubscript{class} = \text{median}_{c=1}^{C} [\text{\emph{WCSR}}(c)]
\end{equation}

$C$ denotes the number of chord classes. $\text{\emph{WCSR}}(c)$ is the WCSR considering only time when chord $c$ is playing. A formal definition of $\text{\emph{WCSR}}(c)$ is also included in Appendix~\ref{app:weighted_chord_symbol_recall_definitions}.

These metrics are intended to measure the model's performance on the long tail of the chord distribution. Measuring both the mean and median is informative as it provides a sense of the skew in performance over classes. While the metric can be defined for any measure of correctness, I report only the \texttt{acc} as I found it to be the most informative. For example, the mean class-wise \texttt{root} score is harder to interpret.

The justification for redefining \texttt{acc}\textsubscript{class} this way is that metrics calculated over discrete frames are not comparable across different frame lengths and are dependent on the method of allocating chords to frames. Instead, continuous measures evaluate models based on the percentage of time that they are correct. This more closely reflect what we truly desire from the model. To illustrate this, imagine a very long frame length. The model could have perfect scores on these frames but be making terrible predictions for much of the song. Through preliminary experiments, it became clear that there are negligible differences between rankings between models in discrete and continuous measures for sufficiently small hop lengths. Nevertheless, I propose that the field of ACR adopts a continuous measure of class-wise accuracy.

I do not also compute \emph{quality}-wise accuracies as computed by \citet{CurriculumLearning}. Quality-wise metrics only ensure that each root is equally weighted. As roots are fairly balanced, this would not add much information. I therefore do not evaluate using quality-wise metrics.

For most experiments, the metrics on the validation set are used to compare performance. The test set is used only to compare the final accuracies of select models in Section~\ref{sec:test-set}.

Other evaluation tools are used such as confusion matrices and the number of chord transitions per song. Note that confusion matrices are calculated using discrete frames for ease of computation. In an ideal setting, these would also be calculated using continuous measures. Given the small differences between the two for short frame lengths, I decided it was not worth the additional engineering effort and computational cost.

\section{Training}\label{sec:training}

For the majority of experiments, I use a random 60/20/20\% training/validation/test over songs in the dataset. This split is kept constant across experiments. This contrasts much of the literature, which uses a 5-fold cross-validation introduced by \citet{FourTimelyInsights}. I did not maintain this status quo in order to obtain clean estimators of the generalisation error using the held-out test set and to save on computation time. For final testing, models are re-trained on the combined training and validation sets and tested on the test set.

Three variants of the dataset are used for training, validation and testing. For training, an epoch consists of randomly sampling a patch of audio from each song in the training set. The length of this sample is kept as a hyperparameter, set to $10$ seconds for the majority of experiments, based on values provided by \citet{StructuredTraining} and a hyperparameter search found in Section~\ref{sec:model_hyperparameters}. For evaluation, the entire song is used as performance was found to be marginally better. This is later discussed in Section~\ref{sec:crnn_performance_across_context}. When validating mid-way through training, songs are split into patches of the same length as the training patches to save on computation time. Samples in a batch are padded to the maximum length of a sample in the batch and padded frames are ignored for loss and metric calculation.

Experiments are run on two clusters with some further evaluation taking place locally. The first is The University of Edinburgh's ML Teaching Cluster. Here, NVIDIA GPUs are used, mostly GTX Titan Xs (12GB VRAM) and RTX A6000s (48GB VRAM) depending on availability on the cluster. Resources have inconsistent availability. Therefore, some experiments are run on Eddie, The University of Edinburgh's research compute cluster, on CPUs due to the lack of availability of GPUs.

Training code is implemented in \texttt{PyTorch}~\citep{pytorch} with seed set to $0$. Unless stated otherwise, models are trained with the Adam optimiser~\citep{adam} with a learning rate of $0.001$ and pytorch's \texttt{CosineAnnealingLR} scheduler, set to reduce the learning rate to 1/10th of its initial value over the run. Models are trained to minimise the cross entropy loss between the predicted chord and the true chord distribution. I use a batch size of 64 and train for 150 epochs unless stated otherwise. Validation part-way through training is conducted every 5 epochs in order to save time. The model is saved whenever the validation loss improves. Each training run takes approximately 30 minutes of GPU time or 1 hour 30 minutes of CPU time. This can vary up to 10 hours of CPU time for experiments with more expensive computations and larger input.

%% file: chapters/crnn.tex
\chapter{A Convolution Recurrent Neural Network}\label{chap:crnn}

In this chapter, I implement a convolutional recurrent neural network (CRNN) from the literature, train it on the \emph{pop} dataset and compare it to two baselines. I then conduct a thorough analysis of the behaviour and failure modes of the CRNN and provide motivation for improvements. 

\section{The CRNN Model}\label{sec:crnn}

I implement a convolutional recurrent neural network (CRNN) as described in \citet{StructuredTraining}, referred to as the \emph{CRNN}. It remains competitive with state-of-the-art, is often used as a comparative baseline and is fast and easy to train.

The model receives input of size $B \times F$ where $B=216$ is the number of bins in the CQT and $F$ is the number of frames in the song. The input is passed through a layer of batch normalisation~\citep{BatchNorm} before being fed through two convolutional layers with a rectified linear unit (ReLU) after each one. The first convolutional layer has a $5\times 5$ kernel and outputs one channel. It is intended to smooth out noise and spread information across adjacent frames. The second layer has a kernel of size $1\times I$ and outputs 36 channels, intended to collapse the information over all frequencies. The output is passed through a bi-directional gated recurrent unit (GRU)~\citep{GRU}, with hidden size initially set to $256$ and a final fully connected layer with softmax activation. This produces a vector of length $C$ for each frame. The chord with the maximum probability is taken as the model's prediction for each frame.

The authors of the model propose using a second GRU as a decoder before the final fully connected layer, called `CR2'. In brief empirical tests, the results with and without `CR2' were very similar. Therefore, I do not include this in the model. Results are left to Appendix~\ref{app:crnn_with_cr2} as they are neither relevant nor interesting.

\subsection{Hyperparameter Tuning}

To ensure that the training hyperparameters are set to reasonable values, I conduct a grid search over learning rates and learning rate schedulers. This is followed by a random search over model hyperparameters. 

\subsubsection{Learning rates}\label{sec:lr_search}

I perform a grid search over learning rates and learning rate schedulers in the set \texttt{[0.1, 0.01, 0.001, 0.0001]} and \texttt{[cosine, plateau, none]} respectively. \texttt{cosine} is as described in Section~\ref{sec:evaluation} and \texttt{plateau} reduces the learning rate to half its current value when validation loss has not improved for 10 epochs and stops training if it has not improved for 25 epochs.

Detailed metrics are left to Appendix~\ref{app:learning_experiment_results}. To summarise, the best-performing model was found to be with \texttt{lr=0.001}. Performance over different schedulers is very similar. I proceed with a learning rate of $0.001$ and \texttt{cosine} scheduling for the rest of the experiments. Scheduling with \texttt{cosine} is chosen as it provides a consistent learning rate reduction. It empirically was found that \text{plateau} would sometimes only reduce the learning rate in the final epochs. Only saving on validation loss improvement provides the regularisation effect of early stopping regardless of the choice of scheduler.

Of the learning rates tested, the best was found to be $0.001$. If any lower, the model does not converge fast enough. If any higher, large gradient updates cause the validation accuracy to be noisy. Figures supporting this conclusion can be found in Appendix~\ref{app:learning_experiment_results}. These figures also show that the validation loss does not increase after convergence. I conclude that the model does not have a propensity to overfit within $150$ epochs, perhaps due to the random sampling of audio patches during training. Combined with the fact that training is relatively quick and that the model is only saved on improved validation loss, I proceed with $150$ epochs of training without early stopping.

Results from \citet{SGD1} suggest that stochastic gradient descent (SGD) can find better minima with a stable learning rate over many epochs. To test this, I trained a CRNN over $2000$ epochs with a learning rate of $0.001$, the \texttt{cosine} scheduler and momentum set to $0.9$. While the model did converge, it did not perform any better than the models trained with Adam. Results are left to Appendix~\ref{app:long_sgd} for lack of interest.

\subsubsection{Model Hyperparameters}\label{sec:model_hyperparameters}

With this learning rate and learning rate scheduler fixed, I perform a random search over the number of layers in the GRU, the hidden size of the layers in the GRU the training patch segment length, the number of convolutional layers prior to the GRU, the kernel size of these layers and the number of channels outputted by each of these layers. The search is performed by independently and uniformly randomly sampling 50 points over discrete sets of possible hyperparameter values. These sets can be found in Appendix~\ref{app:random_hyperparameter_search_sets}. 

Table~\ref{tab:crnn_hparams} shows a sample of the results. Results across hyperparameters are relatively stable. In general, increased model complexity hurts performance and the differences between models are relatively small. Such small differences indicate that the model is learning something simple where increased model complexity does not help. It also seems that the choice of hyperparameters is somewhat arbitrary. The models do not utilise information from a larger context. Deeper layers with more parameters do not help performance. In fact, the parameters suggested by \citet{StructuredTraining} perform just as well as the best-performing hyperparameter selection found in this random search. I therefore proceed with the same hyperparameters suggested by \citet{StructuredTraining} as default for the remainder of experiments.

\begin{table}[h]
    \centering
    \begin{tabular}{ccccccccccc}
        \toprule
        $L$ & $h$ & $k$ & $c$ & $g$ & $ch$ & acc & root & third & seventh & mirex \\
        \midrule
        23 & 231 & 5 & 1 & 1 & 1 & \textbf{59.8} & 78.2 & 74.7 & \textbf{62.0} & \textbf{79.9} \\
        11 & 150 & 7 & 2 & 1 & 3 & 59.6 & 78.5 & 75.0 & 61.9 & 79.2 \\
        43 & 222 & 11 & 2 & 2 & 2 & 59.5 & \textbf{78.7} & \textbf{75.2} & 61.8 & 78.9 \\
        \ldots & \ldots & \ldots & \ldots & \ldots & \ldots & \ldots & \ldots & \ldots & \ldots & \ldots \\
        34 & 159 & 14 & 4 & 2 & 1 & 56.9 & 75.5 & 72.2 & 59.1 & 77.7 \\
        \bottomrule
    \end{tabular}
    \caption{A subset of \emph{CRNN} model results on the large vocabulary with different hyperparameters. The best results for each metric are in boldface. $L$ is the length of training patches of audio in seconds, $h$ and $g$ are the hidden size and number of layers in the GRU respectively and $k$, $c$ and $ch$ are the kernel sizes, number of layers and number of channels in the CNN respectively. Models are ordered by their `Rank', calculated by adding the model's rank order over each metric and ordering by this total. Results across most hyperparameters are very similar. Comparing with the best results from the learning rate search in Table~\ref{tab:crnn_lr}, it seems that the parameters suggested by \citet{StructuredTraining} are good choices. Models with more parameters and longer input tend to perform worse, perhaps due to overfitting. This suggests that the model is learning something simple.}\label{tab:crnn_hparams}
\end{table}


\section{Baseline Models}\label{sec:baselines}

I consider two models as baselines. First, I train a single-layer neural network with softmax activation, which treats each frame of each song independently. The layer receives an input of size $B=216$ and outputs a $C=170$-dimensional vector for each frame. This model is called \emph{logistic} as it can be viewed as a logistic regression model trained using SGD. I could have used a logistic regression model implemented in \texttt{sklearn} but the implementation as a neural network was fast and easy to implement and unlikely to yield significantly different results. Saving on validation loss improvement also provides a regularisation effect.

Secondly, I train a convolutional neural network (CNN). The number of convolutional layers, kernel size and number of channels are left as hyperparameters. The convolutional layers operate on the CQT similarly to how a convolution operates on an image. A ReLU is placed between each layer. These are followed by a 36-channel $1\times I$ convolutional layer and fully connected layer as in the \emph{CRNN}. 

I test models of increasing depths, kernel sizes and channels. In general, the deeper models perform better. Two of these models serve as baselines in reported results. The first model has a single layer and channel and a kernel size of $5$. It serves as an ablation on the GRU part of the \emph{CRNN}. This configuration is referred to as \emph{CNN1}. A second model with 5 layers of kernel size 9, each with 10 channels, is referred to as \emph{CNN5}.

I perform a grid search over learning rates and schedulers for these baselines to ensure that convergence is reached. Convergence results are not meaningfully different than those obtained with the \emph{CRNN} and are hence omitted. I use the best-performing results in each case. This was with a learning rate of $0.001$ for both models and with schedulers of \texttt{plateau} and \texttt{cosine} for \emph{logistic} and \emph{CNN1}/\emph{CNN5} respectively.



\section{First Results}

Table~\ref{tab:first_results} shows the results of the \emph{CRNN} compared with the baseline models. The \emph{CRNN} performs the best out of these models. The GRU layer improves accuracy by $5.2\%$. However, similar performance increases can be achieved by adding convolutional layers as in \emph{CNN5} as opposed to an RNN. Combined with the lack of performance improvement from increasing the audio patch length observed in Section~\ref{sec:model_hyperparameters}, there is strong evidence that the model does not share information across time very far.

We also observe diminishing performance increases with increased model complexity. Performance begins to level out with accuracies of roughly $60\%$. Indeed, the best models trained by \citet{BTC} and by \citet{ChordFormer} never achieve an accuracy of more than $66\%$. \citet{FourTimelyInsights} refer to this as the `glass ceiling' which the field of ACR is still struggling to break through. The problem posed by ACR remains far from solved.

\citet{FeatureMaps} train a deep CNN which remains competitive with state-of-the-art to this day. It contains 8 layers. \citet{BTC} find that the performance of this deep CNN is very similar to that of the \emph{CRNN}, both reaching accuracies of around $65\%$. Training much deeper convolutional networks was found to be more computationally expensive than training the \emph{CRNN} with little performance gain to be had. Therefore, I proceed with the \emph{CRNN} for further experiments.

\begin{table}[h]
    \centering
    \begin{tabular}{lccccccc}
        \toprule
        model & acc & root & third & seventh & mirex & acc\textsubscript{class} & median\textsubscript{class} \\  
        \midrule
        \emph{logistic} & 43.0 & 64.5 & 56.9 & 44.7 & 60.9 & 12.0 & 1.7 \\
        \emph{CNN1} & 54.5 & 74.4 & 69.0 & 56.6 & 73.5 & 16.0 & 2.3 \\
        \emph{CNN5} & 57.8 & 78.1 & 74.0 & 60.0 & 77.8 & 19.2 & \textbf{3.2} \\
        \emph{CRNN} & \textbf{59.7} & \textbf{78.3} & \textbf{75.0} & \textbf{62.0} & \textbf{79.8} & \textbf{19.6} & 2.3 \\
        \bottomrule
    \end{tabular}
    \caption{Results for \emph{logistic}, \emph{CNN1}, \emph{CNN5} and \emph{CRNN}. We see that the \emph{CRNN} performs the best on nearly all metrics. The \emph{CNN5} performs almost as well. This suggests that shallower CNNs can reach similar performance as the deep CNN trained by \citet{FeatureMaps}.}\label{tab:first_results}
\end{table}

\section{Model Analysis}\label{sec:crnn_analysis}

While quantitative metrics summarise how well a model performs over songs, they do not tell us much about the predictions the model makes and where it goes wrong. In this section, I seek to understand behaviour of the model by answering a series of questions.

\subsection{Qualities and Roots}

\textbf{How does the model deal with imbalanced chord distribution?} The class-wise metrics in Table~\ref{tab:first_results} give strong indication that the performance is poor. I use a confusion matrix over qualities of chords to provide more granular detail. 

\begin{figure}[h]
    \centering
    \includegraphics[width=0.8\textwidth]{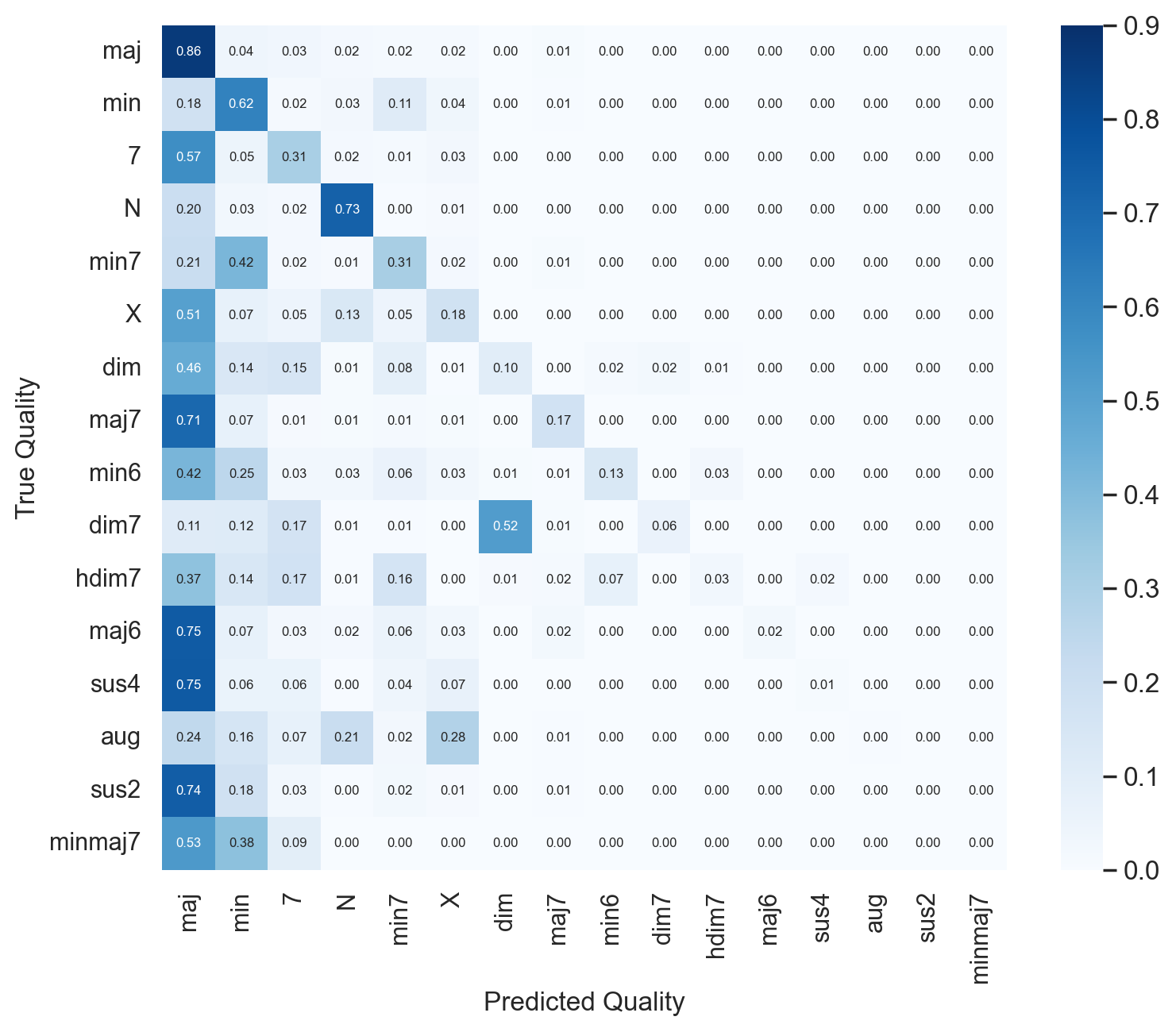}
    \caption{Row-normalised confusion matrices over qualities of the \emph{CRNN} model. Rows are ordered by frequency of chord quality. We observe that the model struggles with the imbalanced distribution. It frequently confuses \texttt{dim7} and \texttt{dim} qualities, consistently predicts \texttt{maj} for \texttt{sus2, sus4, maj6, maj7} and struggles with the rare qualities like \texttt{minmaj7} and \texttt{aug}.}\label{fig:crnn_qual_cm}
\end{figure}

The confusion matrix is illustrated in Figure~\ref{fig:crnn_qual_cm}. The model struggles with rarer chords. On the rarest quality of majorminor7, the model has a recall of $0$. Recall is $0.86$ on the major chord but the model consistently predicts major for similar chord qualities like major7 and sus4. A similar effect is observed with minor chords and qualities like minor7. The model frequently confuses diminished 7 chords for diminished chords. This explains the median class-wise accuracies of nearly $0$ for all models.

I also produce a confusion matrix over roots. This is left to Appendix~\ref{app:cm_roots} as it is less insightful. The model performs similarly over all roots with a recall of between $0.73$ and $0.81$. This is not surprising as the distribution over roots is relatively uniform. Recall on the no chord symbol \texttt{N} is $0.73$. Many of the \texttt{N} chords are at the beginning and end of the piece. The model may struggle with understanding when the music begins and ends. An example of the model erroneously predicting that chords are playing part-way through a song is discussed in Section~\ref{sec:crnn_examples}.

Performance is worse on the unknown chord symbol with a recall of $0.18$. The low performance on \texttt{X} is to be expected. It is a highly ambiguous class with many completely different sounds mapped to it. All of the chords mapped to \texttt{X} will share many notes with at least one class in the known portion of the vocabulary. It is therefore unreasonable to expect the model to be able to predict this class well. This supports the case for ignoring this class during evaluation as is standard in the literature.

\subsection{Transition Frames}\label{sec:transition_frames}

\textbf{Are predictions worse on frames where the chord changes?} Such \emph{transition frames} are present because frames are calculated based on hop length irrespective of the tempo and time signature of the song. Thus, some frames will contain a chord transition. 

To test this, I compute accuracies for transition and non-transition frames separately. The model achieves only $37\%$ on the transition frames compared with $61\%$ on non-transition frames. Therefore, the model is certainly worse at predicting chords on transition frames. Nonetheless, the \emph{CRNN} achieves an overall accuracy of $60\%$. This is because only $4.4\%$ of frames are transition frames with a hop length of $4096$. Improving performance on these frames to the level of non-transition frames would increase the overall frame-wise accuracy by at most $1\%$. 

Through qualitative evaluation discussed in Section~\ref{sec:crnn_examples}, the model was found to struggle with identifying the boundary of a chord change on some songs. This would not be captured by the above metrics if the boundary is ambiguous enough to span multiple frames. Thus, there may be a larger impact in accuracy than a single frame. Furthermore, the ambiguity of chord transition timing will vary over songs. For some songs, this may be the main limiting factor in performance.

\subsection{Smoothness}\label{sec:smoothness}

\textbf{Are the models outputs smooth?} There are over 10 frames per second. If the model outputs rapid fluctuations in chord probability, it will over-predict chord transitions. I use two crude measures of smoothness to answer this question.

Firstly, I look at the number and length of \emph{incorrect regions}. Such a region is defined as a sequence of incorrectly predicted frames with the same prediction. $26.7\%$ of all incorrect regions are one frame wide and $3.7\%$ of incorrect frames have different predictions on either side. This suggests that at least $3.7\%$ of errors are caused by rapidly changing chord predictions. A histogram over incorrect region lengths can be found in Appendix~\ref{app:histogram_over_region_lengths}. This plot shows that the distribution of lengths of incorrect regions is long-tailed, with the vast majority very short.

Secondly, I compare the mean number of chord transitions per song predicted by the model with the true number of transitions per song in the validation set. The model predicts $168$ transitions per song while the true number is $104$. This is convincing evidence that smoothing the outputs of the model could help. 

With these two observations combined, I conclude that further work on the model to improve the smoothness would might performance. Although we might hope to improve on at least $3.8\%$ of errors, this would not improve overall accuracy very much. While rapid changes may be smoothed out, there is no guarantee that smoothing will result in correct predictions. Indeed, it may even render some previously correct predictions erroneous. Nonetheless, the model predicts too many chord transitions. When being used by a musician or researcher, smoothed predictions would be valuable in making the chords more interpretable.

\subsection{Performance Across Context}\label{sec:crnn_performance_across_context}

\textbf{How much does the model rely on context?} I hypothesise that the model is worse at predicting chords at the beginning and end of a patch of audio as it has less contextual information close to these frames. 

To test this, I evaluate the model using the same fixed-length validation conducted during training as described in Section~\ref{sec:training}. Average frame-wise accuracies over the context are then calculated. A plot can be found in Appendix~\ref{app:accuracy_over_context}. I use a segment length of $10$ seconds corresponding to $L=107$ frames. We observe that performance is worst at the beginning and end of the patch but not by much. Performance only dips by $0.05$ at either extreme, perhaps because the model still does have significant context on one side. We can also see that performance starts decreasing 5 or 6 frames from either end, suggesting that this is the extent to which bidirectional context is helpful.

I conduct a further experiment measuring overall accuracy with increasing segment lengths used during evaluation. Results can be found in Appendix~\ref{app:accuracy_vs_context_length}. The plots show that accuracy increases by $0.5$ after increasing the segment length from 5 seconds to 60 seconds. Although this is not much of an increase, it confirms that it is better to evaluate over the entire song at once.

\subsection{Consistency Over Songs}

\begin{figure}[ht]
    \centering
    \includegraphics[width=1.0\textwidth]{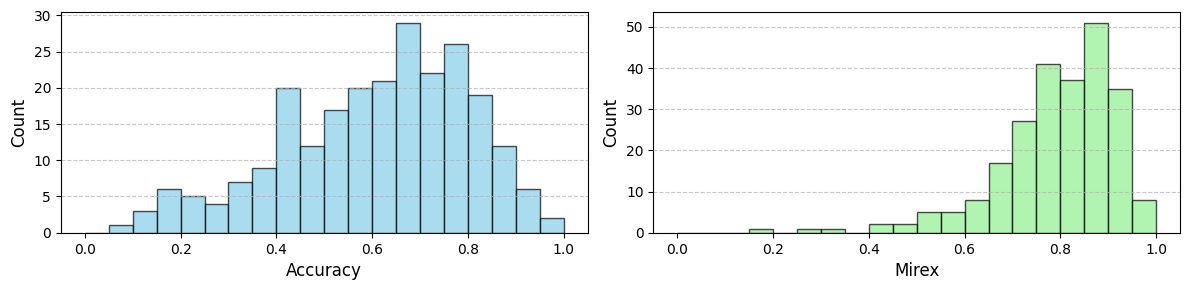}
    \caption{Histogram of accuracies and mirex scores over songs in the validation set. Accuracies are mixed, with $15\%$ of songs below $40\%$, and $69\%$ between $0.4$ and $0.8$. However, with the more generous \texttt{mirex} metric, we find that there are almost no songs below a score of $40\%$ and only $7\%$ below $0.6$. This suggests that many of the mistakes the model makes are small, like predicting \texttt{C:maj} instead of \texttt{C:maj7}. The very low outliers in the \texttt{mirex} score were found to be songs with incorrect annotations found in Section~\ref{sec:data-integrity}.}\label{fig:crnn_song_hist}
\end{figure}

\textbf{Does the model have consistent performance over different songs?} The set of accuracies over songs of the \emph{CRNN} has a standard deviation of $13.5$. This suggests that performance is not stable over songs. To provide further insight, I plot a histogram of accuracies and \texttt{mirex} scores over the validation set in Figure~\ref{fig:crnn_song_hist}. We observe that the model has mixed performance with accuracy, with $15\%$ of songs scoring below $40\%$.  

When we use the more generous \texttt{mirex} metric, there are very few songs below $40\%$ and only $7\%$ are below $0.6$. This large discrepancy between accuracy and \texttt{mirex} suggests that many of the mistakes that the model makes are small. These mistakes are a good guess in the sense that the prediction may have omitted a seventh or mistaken a major 7 for its relative minor. Examples of such mistakes are discussed in Section~\ref{sec:crnn_examples}. 

I conclude that many of the model's predictions are reasonable but often lack the detail contained in good annotations like correct upper extensions. Whether these reasonable guesses are correct can vary widely over songs.

\subsection{Four Illustrative Examples}\label{sec:crnn_examples}

\begin{figure}[ht]
    \centering
    \includegraphics[width=1.0\textwidth]{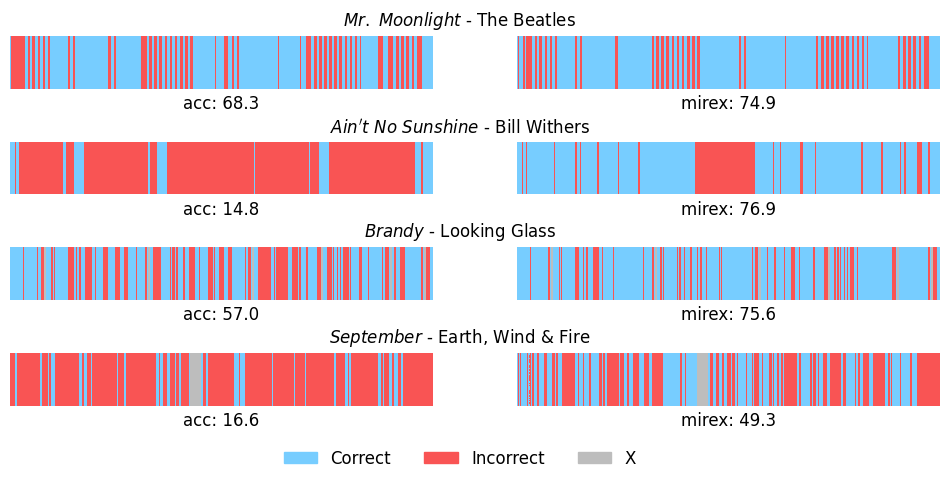}
    \caption{Chord predictions of the \emph{CRNN} model on four songs from the validation set (blue: correct, red: incorrect, gray: \texttt{X}). This allows us to understand some of the behaviour of the model. We can see regular, repeated errors in \emph{Mr.\ Moonlight}, which are mostly mistaking two similar qualities. The discrepancy between accuracy and \texttt{mirex} on \emph{Ain't No Sunshine} can be explained by missing sevenths in many predictions. The large incorrect region is a voice and drum only section where the model continues to predict chords due to implied harmony by the melody. Predictions in \emph{Brandy} are quite good in general, though many errors arise from incorrectly predicting the boundaries of chord changes. The model struggles with \emph{September}, missing chord boundaries and sometimes predicting completely wrong chords. There are clearly songs where the model's outputs are less sensible. However, most of the model's mistakes can be explained and are reasonable.}\label{fig:crnn_examples}
\end{figure}

Now, let us inspect predictions for a few songs to see how the model performs. I choose four examples showing different behaviours and failure modes of the model. Predictions are illustrated frame-by-frame and coloured by correctness, as measured by both accuracy and \texttt{mirex} score in Figure~\ref{fig:crnn_examples}. 

In \emph{Mr.\ Moonlight}, there are few differences between the accuracy and \texttt{mirex} score. There are regular, repeated errors, many of which are mistaking \texttt{F:sus2} for \texttt{F:maj}. This is an understandable mistake to make, especially after hearing the song and looking at the annotation where the main guitar riff rapidly alternates between \texttt{F:maj} and \texttt{F:sus2}. The confusion matrix in Figure~\ref{fig:crnn_qual_cm} suggests this mistake is very fairly common on qualities like \texttt{sus2} which are similar to \texttt{maj}. 

In \emph{Ain't No Sunshine}, the \texttt{mirex} score is significantly higher than the accuracy. This is because the majority of the mistakes the model makes are missing a seventh. For example, the model predicts \texttt{A:min7} for the true label of \texttt{A:min7} or \texttt{G:maj} for \texttt{G:7}. Other mistakes that \texttt{mirex} allows for include confusing the relative minor or major such as predicting \texttt{E:min7} when the chord is \texttt{G:maj}. All of these mistakes occur frequently in this song. The mean difference between the accuracy and mirex is $18.7\%$, with one song reaching a difference of over $70\%$. Hence, we can attribute many of the model's mistakes to such behaviour. `Ain't no Sunshine' also contains a long incorrect section in the middle. This is a section with only voice and drums, which the annotation interprets as \texttt{N} symbols, but the model continues to predict harmonic content. The model guesses \texttt{A:min} throughout this section. This is a sensible label as when this melody is sung elsewhere in the song, it is labelled as \texttt{A:min7}.

In the next two songs, \emph{Brandy} and \emph{September}, the model's mistakes are less interpretable. While performance is okay on \emph{Brandy} with a \texttt{mirex} of $75.6\%$, the model struggles with the boundaries of chord changes, resulting in sporadic short incorrect regions in the figure. In `Earth, Wind and Fire', the model struggles with the boundaries of chord changes and also sometimes predicts completely wrong chords which are harder to explain. Listening to the song and inspecting the annotation makes it apparent that this is a difficult song for even a human to annotate well, and similarly, the model does not fare well.

\section{Takeaways and Next Steps}

Below, I summarise the main takeaways from this section and motivate further improvements to the model.

\textbf{Performance on rare chord classes is poor.} There are few instances of chord classes with complex qualities and upper extensions. The model ends up predicting major and minor classes for these rare chords. There are many methods of addressing an imbalanced distribution in machine learning. The simplest is to add a weighting to the loss function which I explore in Section~\ref{sec:weighted_loss}. I also look at a `structured' loss function which exploits similarity between chords in Section~\ref{sec:structured_loss}. Performance might also be improved through better data. I explore the use of data augmentation in Section~\ref{sec:pitch-augmentation} and synthetic data generation in Section~\ref{sec:synthetic_data}.

\textbf{Predictions are not smooth.} While it is unclear whether or not smoothness will improve performance, a good chord recognition model's predictions would be smooth. Musicians do not expect chords to change every $93$ms. This motivates the exploration of a `decoding' step in Section~\ref{sec:decoding}.

\textbf{The model does not use long-range context.} \emph{CNN1} only shares context a maximum of $5$ frames either side as this is its kernel size. It achieves an accuracy of $54.5\%$, just $5\%$ less than the \emph{CRNN}. This suggests that most of the performance gain associated with including contextual information is neither complex nor far-reaching. I conclude that while a little context improves performance, the \emph{CRNN} does not use context in a complex manner.


\textbf{The model is simple.} The analysis of feature maps by \citet{FeatureMaps} corroborate up with this idea. Their analysis suggests that the model detects the presence of individual notes and decides which chord is present based on these notes. This is why more parameters do not help. Unfortunately, this results in many similar chords being confused. The root note can be wrong. Similar qualities are often mistaken. Predictions often miss upper extensions. This offers an explanation for the large discrepancy between accuracy and \texttt{mirex} score with average values of $60\%$ and $79\%$ over the validation set, respectively. This problem is exacerbated by the imbalance in the dataset discouraging the model from being sensitive to indicators of rare chord classes.

\textbf{Performance is song-dependent.} Accuracies over songs vary widely. \texttt{mirex} scores are more consistent but still vary. A detailed analysis of the properties of songs with poorer performance would be valuable work. I will not explore this further here beyond further qualitative analysis.

\textbf{Chords are not interpretable in time.} The model struggles a little on transition frames. Solely improving performance on such frames is unlikely to improve metrics by much. A much better reason to segment chords is to give the output of the model a far more interpretable meaning. The frame-wise correctness plots illustrated in Figure~\ref{fig:crnn_examples} are not musically interpretable. Even if chord symbols were added, this would not constitute good musical notation. Musicians do not operate over $93$ms frames. They think of music as existing in beat space. \citet{AlignmentChordAnnotations} explore the related task of finding chord boundaries in audio given the chord sequence. Instead, I take inspiration from \citet{MelodyTranscriptionViaGenerativePreTraining} and use a beat detection model to task the model with predicting chord over beats rather than frames in Section~\ref{sec:beat-synchronisation}. 

%% file: chapters/model_improvements.tex
\chapter{Improving the Model}\label{chap:model_improvements}

In this chapter, I use the insights from Chapter~\ref{chap:crnn} to improve the \emph{CRNN} and address questions raised by the literature. I perform a series of experiments to test improvements to the model, evaluate a selection of models on the test set and perform a qualitative analysis of the model's outputs.

Many of these experiments introduce new hyperparameters. I choose these hyperparameters in a greedy fashion and keep them as specified unless stated otherwise. While the assumption of independence between hyperparameters is undoubtedly wrong, performing a full hyperparameter search is computationally infeasible.

I first conduct experiments verifying that CQTs are the best features for ACR and that a hop length of $4096$ is appropriate. These experiments are detailed in Appendix~\ref{sec:spectrogram-results}. To summarise, CQTs achieve $10\%$ greater accuracy than other spectrogram variants and any hop length less than $4096$ achieves similar results. Thus, I proceed with CQTs and a hop length of $4096$.

\section{Decoding}\label{sec:decoding}

As observed in~\ref{sec:smoothness}, the \emph{CRNN} predicts $168$ transitions per song as opposed to the $104$ seen in the ground truth data. I implement a decoding step over the frame-wise probability vectors to smooth predicted labels. Common choices for decoding models include a conditional random field (CRF)~\citep{ACRLargeVocab1, BTC} and a hidden Markov model (HMM)~\citep{BalanceRandomForestACR}. 

I first implement a HMM. The HMM treats the frame-wise probabilities as emission probabilities and the chord labels as hidden states. \citet{CQTvsChroma} note that using a transition matrix with homogeneous off-diagonal entries in the transition matrix performs similarly to using a learned transition matrix. I adopt such a transition matrix for this HMM, with a parameter $\beta$ denoting the probability of self-transition and all other transition probabilities equal to $\frac{1-\beta}{C-1}$. Decoding then follows the Viterbi algorithm~\citep{Viterbi} over the summed forward and backward pass.

\begin{figure}[ht]
    \centering
    \includegraphics[width=0.8\textwidth]{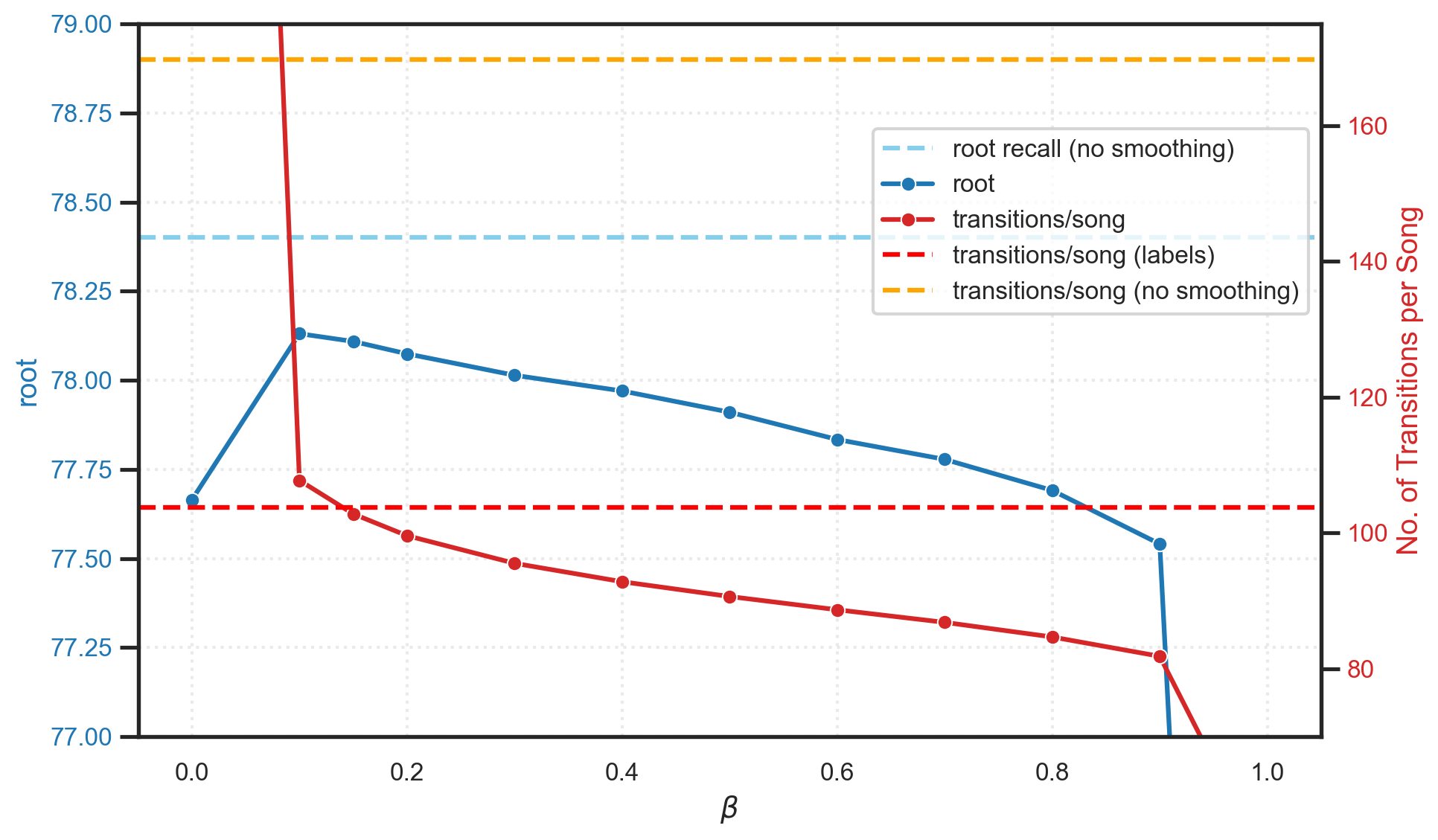}
    \caption{Effect of the HMM smoothing parameter $\beta$ on the \emph{CRNN} model. As we increase $\beta$, the number of transitions per song decreases. I choose $\beta = 0.15$ as it results in $102$ transitions per song, very close to the $104$ of the ground truth. Performance is stable across $\beta$ with a slight degradation for $\beta > 0.3$. Other performance metrics showed similarly stable results. }\label{fig:hmm_beta_search}
\end{figure}

A plot of the effect of $\beta$ on the model's performance and the number of transitions per song is shown in Figure~\ref{fig:hmm_beta_search}. From this plot, we conclude that smoothing has little effect on the performance of the model while successfully reducing the number of transitions per song to that of the true labels. I choose $\beta = 0.15$ for the remainder of experiments as it results in $102$ transitions per song while maintaining high performance. 

The effect of the HMM on the incorrect regions previously discussed in Section~\ref{sec:smoothness} can be found in Appendix~\ref{app:histogram_over_region_lengths}. The HMM reduced the percentage of incorrect regions which are a single frame long from $26.7\%$ to $16.7\%$. A more intuitive way to see the effect of the HMM is to look at a section of a song for which the model previously predicted many chord transitions for. This is illustrated in Appendix~\ref{app:hmm_smoothing_effect}.

I also implement a linear chain CRF using \texttt{pytorch-crf}.\footnote{\url{https://github.com/kmkurn/pytorch-crf}} In contrast to the HMM, the CRF uses a learned transition matrix. Results comparing the HMM, CRF and no smoothing can be found in Table~\ref{tab:smoothers}. Both the CRF and HMM reduce the number of transitions per song to a similar level. The HMM outperforms the CRF with $3.5\%$ greater accuracy. The HMM has almost identical performance to the model with no smoothing. I hypothesise that the learned transition matrix allows the model to overfit to the chord sequences in the training set. Regardless of the explanation, I proceed with HMM smoothing.

\begin{table}[ht]
    \centering
    \begin{tabular}{lcccccc}
        \toprule
        smoother & acc & root & third & seventh & mirex & transitions/song\\  
        \midrule
        none & \textbf{60.0} & \textbf{78.1} & \textbf{75.0} & \textbf{62.3} & \textbf{79.2} & 167 \\
        HMM & \textbf{60.0} &\textbf{78.1} & \textbf{75.0} & \textbf{62.3} & \textbf{79.2} & \textbf{102} \\
        CRF & 56.5 & 75.5 & 72.5 & 58.6 & 76.2 & 100 \\
        \bottomrule
    \end{tabular}
    \caption{Results for the HMM and CRF smoothing methods with the \emph{CRNN}. The HMM has almost identical performance to the model with no smoothing. Smoothing must correct predictions for roughly the same amount of time as it introduces errors. However, it drastically reduces the number of transitions per song to an acceptable level. The CRF performs notably worse than the HMM. }\label{tab:smoothers}
\end{table}

\section{The Loss Function}

\subsection{Weighted Loss}\label{sec:weighted_loss}

One of the most significant with the \emph{CRNN} is the low recall on rarer chord qualities. Two standard methods for dealing with long-tailed distributions are weighting the loss function and over-sampling. \citet{CurriculumLearning} also explore the use of curriculum learning as a form of re-sampling which we do not explore here because they report only minor performance gains. Sampling is explored by \citet{BalanceRandomForestACR}, but they use a different model based on pre-computing chroma vectors and re-sampling these chroma vectors for use in training a random forest for frame-wise decoding. 

In our setting, re-sampling training patches of audio may be interesting to explore but is left as future work. It would require a complex sampling scheme as frames cannot be sampled independently. 

Weighting has been explored by \citet{ACRLargeVocab1}. We employ a similar but simpler implementation here. A standard method of weighting is to multiply the loss function by the inverse frequency of the class of the current training sample, with a parameter controlling the strength of the weighting. This is defined in Equation~\ref{eq:weighting}.

\begin{equation}\label{eq:weighting}
    w_c = \frac{1}{{(\text{count}(c) + 10)}^\alpha}
\end{equation}

Where $w_c$ is the weight for chord $c$, $\text{count}(i)$ is the number of frames with chord $c$ in the dataset and $\alpha$ is a hyperparameter controlling the strength of weighting. $\alpha=0$ results in no weighting and increasing $alpha$ increases the severity of weighting. I add $10$ in the denominator to avoid dividing by $0$ and to diminish the dominating effect of chords with very few occurrences. I then define normalised weights $w_c^*$ in Equation~\ref{eq:weighted_loss} so that the learning rate can remain the same.

\begin{equation}\label{eq:weighted_loss}
    w_c^* = \frac{w_c}{s} \text{ where } s = \frac{\sum_{c\in \mathcal{C}} \text{count}(c)\cdot w_c}{\sum_{c\in \mathcal{C}} \text{count}(c)}
\end{equation}

\begin{figure}[H]
    \centering
    \includegraphics[width=0.8\textwidth]{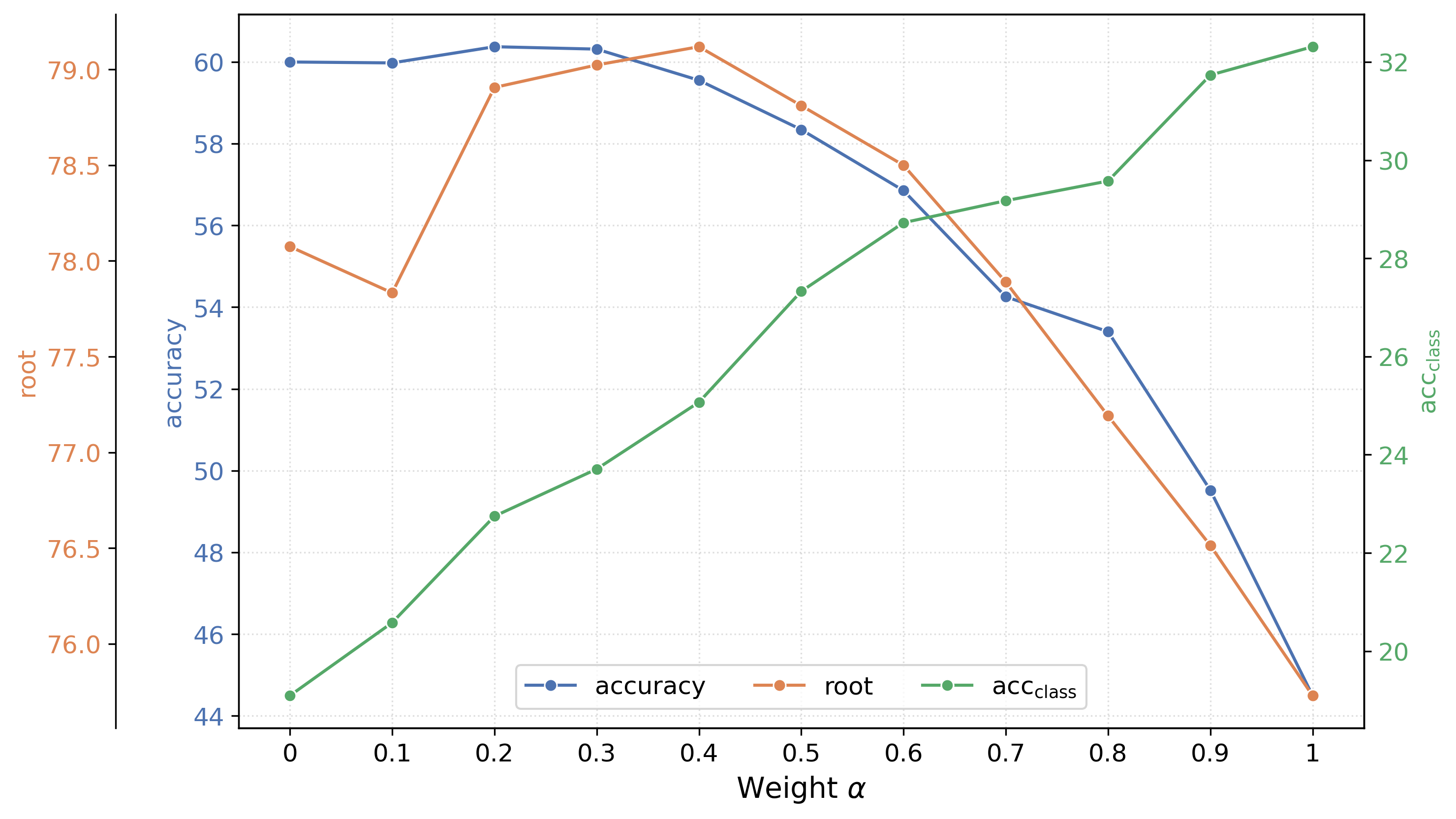}
    \caption{Effect of weighted loss on the \emph{CRNN} model with varying $\alpha$. As we increase $\alpha$, \texttt{acc}\textsubscript{class} improves accuracy and \texttt{root} decreases. I claim there is a sweet spot where very little overall performance is sacrificed for better class-wise accuracies. I choose this to be $\alpha = 0.3$. The \texttt{root} and \texttt{third} metrics improve and less than $3\%$ is lost on other metrics while mean class-wise accuracy improves by $6\%$.}\label{fig:weighted_loss}
\end{figure}

Where $\mathcal{C}$ is the set of all chords in the vocabulary. This keeps the expected weight over samples at $1$ such that the effective learning rate remains the same. These values are calculated over the training set. I test values of $\alpha$ in the set \{0, 0.05, 0.1, \ldots, 0.95, 1\}. The plot in Figure~\ref{fig:weighted_loss} illustrates the effect of the weighting on the model's performance. I find that increasing $\alpha$ improves \texttt{acc}\textsubscript{class} but decreases \texttt{root} accuracy. Choosing $\alpha = 0.3$ maximises \texttt{acc}\textsubscript{class} without hurting root accuracy which I carry forward to subsequent experiments.

For further insight, a plot of the differences between confusion matrices with and without weighted loss can be found in Appendix~\ref{app:weighted_loss_confusion_matrix}. Notably, recall on most qualities increases, with recall on major7 doubling to $0.34$. The weighted model predicts $2.2$ times fewer \texttt{X} symbols, which may explain how it increases recall on these rarer qualities without sacrificing accuracy. 

Weighting the loss function also slightly increases the number of transitions predicted per song. This may be because occasional sharp gradient updates cause more extreme probability outputs. I increase the HMM smoothing parameter $\beta$ to $0.2$ to bring the number of transitions per song to $104$.

\subsection{Structured Loss}\label{sec:structured_loss}

\citet{StructuredTraining} propose a structured loss function, which they claim improves performance on the \emph{CRNN} model. They introduced additional targets for the root, bass and pitch classes. I follow a similar method but do not include the bass as the current chord vocabulary does not consider inversions. The idea behind this loss term is to explicitly task the model with identifying the components of a chord we care about. This can allow the model to exploit structure in the chord vocabulary such as shared roots and pitch classes, rather than all symbols being predicted independently.

The root can be any of the 12 notes in the Western chromatic scale, \texttt{N} or \texttt{X}, creating a 14-dimensional classification problem. The 12 pitch classes each represent a single binary classification problem. Two fully connected layers calculate a 14-dimensional vector and 12-dimensional vector from the hidden representation outputted from the GRU for the root and pitch classes, respectively. Finally, these representations are concatenated with the GRU representations and fed into the final fully connected layer to predict the chord symbol.

The mean cross-entropy loss is calculated in each case. These are summed to form the \emph{structured loss}. Finally, a linear combination of the structured and original losses is calculated. The final loss is a convex combination of the original loss and the structured loss as defined in Equation~\ref{eq:structured_loss}.

\begin{equation}\label{eq:structured_loss}
    L = \gamma L_{chord} + (1-\gamma)(L_{root} + L_{pitch})
\end{equation}

Where $L$ is the overall loss, $L_{chord}$ is the cross-entropy loss over chords symbols, $L_{root}$ is the cross-entropy loss targeting the root, $L_{pitch}$ is mean binary cross-entropy over each of the pitch classes. $\gamma$ is a hyperparameter controlling the weighting of the original loss. 

I test models with $\gamma\in\{0, 0.1 \ldots 0.9\}$. Choosing $\gamma=0.7$ improves accuracy by $1.3\%$ while \texttt{mirex} worsens by $0.3\%$. Accuracy with \texttt{third} increases by $1.7\%$ and on \texttt{seventh} by $1.3\%$. Generally, greater $\gamma$ improves accuracy metrics while \texttt{mirex} results are noisy. A plot of the trend can be found in Appendix~\ref{fig:structured_loss} but does not provide further insight. I keep $\gamma=0.7$ from now on based on peak accuracy. 

\section{Generative Features}\label{sec:generative_features}

\citet{MelodyTranscriptionViaGenerativePreTraining} use generative features extracted from Jukebox~\citep{Jukebox} to improve performance for melody transcription. They also produce a chord transcription model using the same methodology but do not report results. I decide to test generative features using MusicGen~\citep{MusicGen} as a feature extractor. This was for several reasons. MusicGen is a newer model. It has several different sizes of models that can be tested against each other as an experiment on the complexity of the model. It has a fine-tuned variant called MusiConGen~\citep{MusiConGen} which is used for synthetic data generation in Section~\ref{sec:synthetic_data}. All model weights are available on the HuggingFace Hub.\footnote{\url{https://huggingface.co/docs/hub/en/index}} Finally, its training data was properly licensed, unlike Jukebox. I leave the results of Jukebox for future work. Details of the feature extraction process are left to Appendix~\ref{app:generative_feature_extraction}.

As the representations are $2048$-dimensional, it is computationally infeasible to feed this directly into the GRU. Instead, using fully connected layers, I project these vectors down to a power of $2$, from $16$ to $1024$. The best representation is with $64$ dimensions, although results show no clear trend. Results are left to Appendix~\ref{app:projection_dimensionality}.

I test different variants of MusicGen, including musicgen-large (3.3B parameters), musicgen-small (330M), musicgen-melody (1.5B) and MusiConGen (1.5B). I also test different reductions of the four `codebooks' outputted by the language model. This includes taking each codebook on its own, the element-wise mean across codebooks and the concatenation of all four. For further details on codebooks, see the work of \citet{MusicGen}. The best model was found to be musicgen-large and the best reduction to average over the four codebooks. However, results were close and are left to Appendices \ref{app:generative_feature_extraction_models} and \ref{app:generative_feature_extraction_reductions} for lack of importance.


Surprisingly, the concatenated representation performs worse than the averaged representation as it contains at least as much information. However, if the information provided by each codebook is essentially the same, then there are no reasons that the concatenated representation should perform better and training with Adam may simply find a worse minimum. Training on $8192$-dimensional vectors is also computationally expensive.

To test whether or not these features help when compared with a CQT, I test with the CQT only, generative features only and a concatenation of the two. The results are shown in Table~\ref{tab:gen_feature_comparison}. Although the generative features perform worse than the CQT, they contain information useful for chord recognition with an accuracy of $58.7\%$. Performance remains largely the same when the CQT and generative features are used together. This experiment was run multiple times, with similar results each time. There is no clear evidence that the generative features provide any benefit over just using the CQT. 

This conclusion is surprising as \citet{MelodyTranscriptionViaGenerativePreTraining} claim that generative features are better than hand-crafted features for the related task of melody recognition. However, they only compare to mel-spectrograms, which may not perform as well as CQTs, as they certainly do not for chord recognition. Observations here cast doubt on how well their claims generalise to chord recognition. Features extracted from other generative models such as Jukebox~\citep{Jukebox} or MusicLM~\citep{MusicLM} may perform better. The comparison is left for future work.

Given the lack of improvement and the drastically increased computational cost associated with extracting features and training the model, I do not proceed with training on generative features.

\begin{table}
    \centering
    \begin{tabular}{lccccc}
        \toprule
        features & accuracy & root  & third & seventh & mirex \\  
        \midrule
        gen$\cdot$CQT  & \textbf{61.0}     & \textbf{80.3}  & \textbf{77.0}  & \textbf{63.3}    & 78.4  \\
        gen      & 58.7     & 77.6  & 74.3  & 60.9    & 77.5  \\
        CQT      & \textbf{61.0}     & 79.8  & 76.8  & 63.2    & \textbf{79.3}  \\
        \bottomrule
    \end{tabular}
    \caption{Comparison of CQT and generative features with a concatenation of the two, denoted as gen$\cdot$CQT. The concatenation performs the best on most metrics but not by enough to claim is meaningfully better. }\label{tab:gen_feature_comparison}
\end{table}

\section{Pitch Augmentation}\label{sec:pitch-augmentation}

Pitch augmentation has been done in other works on chord recognition, either on the CQT~\citep{ACRLargeVocab1} or on the audio~\citep{BTC,StructuredTraining}. Although similar, these are not identical transformations. Shifting the CQT takes place after discarding phase information and leaves empty bins behind, whereas audio pitch shifting can introduce other artefacts intended to preserve harmonic structure and maintain phase information. I implement both methods and compare them.

\begin{figure}[H]
    \centering
    \includegraphics[width=1.0\textwidth]{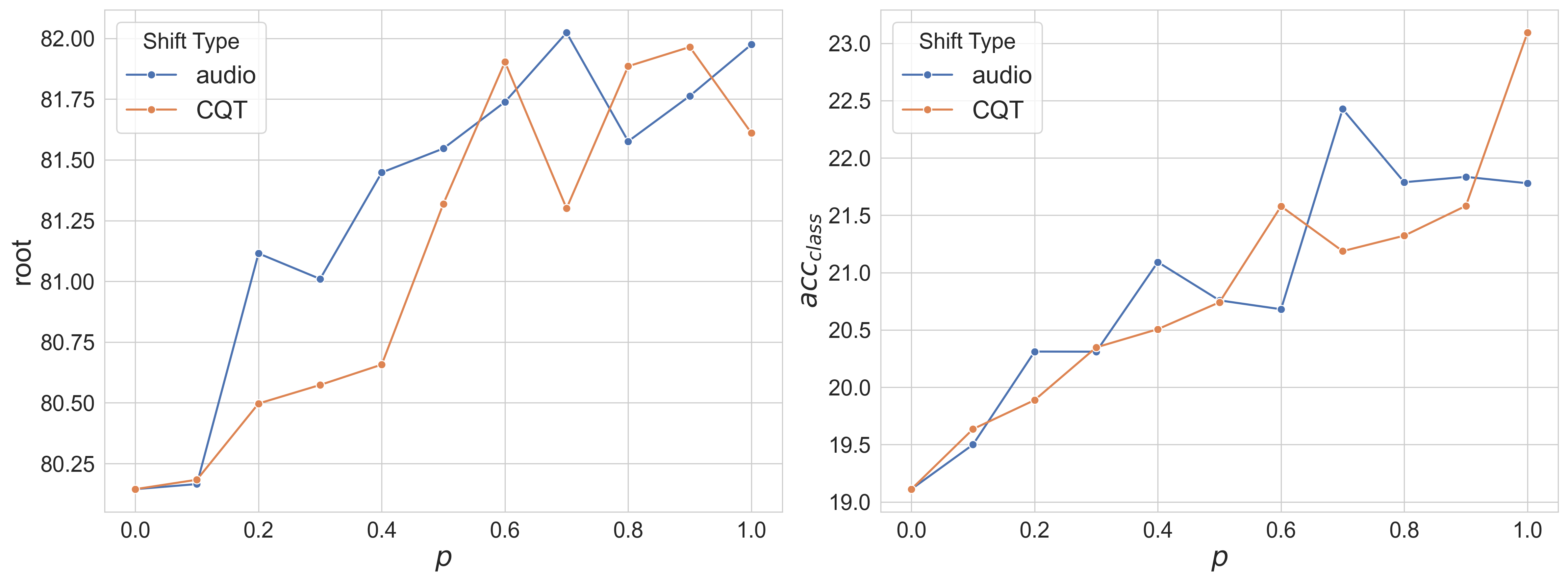}
    \caption{Effect of pitch shift probability $p$ on \texttt{root} (left) and acc\textsubscript{class} (right). Pitch shifting provides a clear increase in performance. Shifting the audio or the CQT has a similar effect. I choose $p=0.9$, which is close to making the model root-invariant.}\label{fig:pitch_augmentation}
\end{figure}

When a sample is drawn from the training set, it is shifted with probability $p$. The shift is measured in semitones in the set $\{-5,-4\ldots -1, 1, \ldots 6\}$ with equal probability of each shift. This results in 12 times as much training data. Convergence is still reached in 150 epochs. Shifting the CQT matrix is done by moving all items up or down by the number of bins corresponding to the number of semitones in the shift. The bins left behind were filled with a value of -80dB. Audio shifting is done with \texttt{pyrubberband}.\footnote{\url{https://github.com/bmcfee/pyrubberband}}. CQTs are then calculated on the shifted audio. A plot of the effect of the shift probability $p$ on the model's performance can be found in Figure~\ref{fig:pitch_augmentation}.

Results show a clear trend that increasing $p$ improves performance. Shifting the audio provides a very similar effect to simply shifting the CQT. Choosing $p=0.9$ results in a $2.1\%$ increase in accuracy. The \texttt{mirex} score breaks the trend with performance varying over different values for $p$. \texttt{acc}\textsubscript{class} also improves by more than $2\%$. This can be explained by the model becoming becoming root-invariant. With $p=0.9$, all potential roots become close to equally likely. I proceed with pitching shifting with $p=0.9$ on the CQT for the remainder of the experiments as it is computationally cheaper than shifting the audio.

\citet{StructuredTraining} claim an increase of $5\%$ on the median across most metrics. I do not find such a large effect here. Nonetheless, pitch shifting is a useful augmentation.

Note that the weights for the weighted loss are calculated based on \emph{expected} counts, taking into account the shift probability $p$. I also test shifting on both the CQT and audio but the results are not different than shifting with either method alone. Unfortunately, it was not computationally feasible to test pitch shifting with generative features as the feature extraction over $12 * 1210 = 14,520$ songs is too expensive.

\section{Synthetic Data Generation}\label{sec:synthetic_data}

Given the success of pitch augmentation for ACR, it is sensible to look for other sources of data. Further augmentation is possible by adding noise and time-stretching. However, these do not provide new harmonic structures or create new instances of rare chords, so I do not explore these here. Instead, I look to generate new data, taking into account our understanding of harmonic structure. Generation would be possible through automatic arrangement, production and synthesis software. However, this is a complex task, requires a lot of human input and is unlikely to produce sufficient variety of timbres, instrumentation and arrangement. Instead, I use a recent chord-conditioned generative model called MusiConGen~\citep{MusiConGen}. I use this over CocoMulla~\citep{CocoMulla} as the method of chord-conditioning has a more straightforward interface, doesn't require reference audio and the authors claim that it adheres more closely to its conditions. Indeed, they feed the outputted audio through BTC~\citep{BTC} and find a \texttt{triads} score of $71\%$ using the chord conditions as the ground truth. While far from perfect, it suggests the model can generate audio that mostly adheres to the chord conditions.

I generate $1210$ songs, each $30$ seconds long, to mimic the size of the \emph{pop} dataset. I refer to this dataset as \emph{synth}. This is split into a train, validation and test split in the same fashion as for the \emph{pop} dataset. Generating a larger order of magnitude of songs would require a lot of compute. While the model supports auto-regressive generation for longer audio, its outputs become incoherent using the provided generation functions. It also sometimes produces incoherent output with $30$ seconds of generation, but this was much less common.

To generate a song, I sample a BPM from a normal distribution with mean $117$ and standard deviation $27$, clipped to lie in the range $[60,220]$. These values were calculated from the training set. I then sample a song description from a set of $20$ generated by ChatGPT. The descriptions outline a genre, mood and instrumentation. Descriptions include only jazz, funk, pop and rock, which were all part of the fine-tuning training set for MusiConGen. The model does not output melodic vocals, owing to the lack of vocal music in the pre-training and fine-tuning data. Finally, I generate a jazz chord progression using the theory of functional harmony~\citet{GenerativeGrammarJazz}. Details of this generation process can be found in Appendix~\ref{app:jazz_chord_progression_generation}. The process generates a very different chord distribution than the one in \emph{pop} with many more instances of upper extensions and rare qualities. This is intended to provide the model with many more instances of rare chords. 

To offset this distribution shift, I calibrate the probabilities outputted by the model. To encourage root-invariance, calibration terms are averaged over roots for each quality. The details of calculating calibration terms are described in Appendix~\ref{app:calibration} with a figure showing the calibration terms for each quality. This figure shows that the rarer qualities are much more common in the synthetic data.

I manually inspected outputs from MusiConGen. In general, the outputs are good. They consistently stick to the provided BPM and usually stick to the chord conditions. Outputs are occasionally musically strange with jarring drum transients and unrealistic chord transitions. They also did not have an enormous variety in timbre and instrumentation. Nonetheless, most examples have sensible annotations that one would expect a human to be able to annotate well. 

I compare a model trained on only \emph{pop}, trained on only synthetic data and trained on both. While the latter results in more training data per epoch, convergence is always reached, so these are fair comparisons. I test the models on the \emph{pop} and synthetic data validation splits. Given the increased instances of rare chords in the synthetic data, I remove the weighting on the loss function for the models trained on synthetic data.

Table~\ref{tab:synthetic_data} shows the results. The model trained on both datasets performs very similarly to the model only trained on \emph{pop}. That is, the model has not overfitted to the synthetic data. However, it has also not improved performance. Training on synthetic data drastically improves the accuracy on the \emph{synth} validation split. This is likely due to the unique chord distribution of the generated jazz progressions and the consistent instrumentation. The model trained on both datasets performs the best, with an accuracy of $51.0\%$.

\begin{table}[h]
    \centering
    \begin{tabular}{lcccc}
        \toprule
        training set & \emph{pop} acc & \emph{pop} third & \emph{pop} seventh & \emph{synth} acc \\  
        \midrule
        \emph{pop} & \textbf{62.1} & 77.7 & 64.4 & 24.2 \\
        \emph{synth} & 48.6 & 64.4 & 50.6 & 44.8 \\
        both & \textbf{62.1} & \textbf{77.9}  & \textbf{64.6} & \textbf{51.0} \\
        \bottomrule
    \end{tabular}
    \caption{Results for models trained on \emph{pop}, \emph{synth} and both together. Unsurprisingly, models trained on \emph{pop} perform better on \emph{pop} and models trained on \emph{synth} perform better on \emph{synth}. The model trained on both does not do meaningfully better on the \emph{pop} validation split, but performance increases on the \emph{synth} validation split. Training only on \emph{pop} results in an accuracy of just $24.2\%$ on \emph{synth}, much lower than reported by \citet{MusiConGen}. However, this may be due to the unrealistic distribution of chords in the generated sequences. }\label{tab:synthetic_data}
\end{table}

For further insight, the difference in confusion matrices over qualities is plotted in Appendix~\ref{app:cm_synthetic_data}. There are few clear trends. Recall increases for some rare qualities and decreases for others. Notably, recall on \texttt{min7} improves by $13\%$ and the model is much better at predicting the third in dominant and minor 7 chords. This is likely why the third and seventh accuracies increase slightly on the \emph{pop} validation set. 

While performance does not meaningfully improve on \emph{pop}, the lack of overfitting and the gain on the synthetic data provide hope that synthetic data could prove a useful tool for ACR with further work and improved generative models. Furthermore, producing hand-annotated data is error-prone and subject to human interpretation. Synthetic data may produce non-identifiable data points but is consistent and error-free. Poor performance on synthetic data can only be explained by failures of the model or indeterminacy of the generated samples, not incorrect labels. Regardless, given the lack of performance increase, I do not continue to train on synthetic data.

\section{Beat Synchronisation}\label{sec:beat-synchronisation}

Chords exist in time. Musicians interpret chords in songs as lasting for a certain number of beats, not a fixed length in time. In its current form, the model outputs frame-wise predictions. While these could be stitched together to produce a predicted chord progression or beat-wise predictions could be made as a post-processing step, I decide to implement a model that outputs beat-wise predictions directly. This allows the model to use information from the entire duration of the beat to make its prediction.

Following the methodology of \citet{MelodyTranscriptionViaGenerativePreTraining}, I first detect beats using \texttt{madmom}~\citep{madmom}. This returns a list of time steps where beats have been detected. I first verify that these beats are plausible. I perform a cross-correlation analysis with the chord transitions similarly to Section~\ref{sec:data-integrity}. A histogram of maximum lags within a window of $0.3$ seconds can be found in Appendix~\ref{app:maximum_lag_cross_correlation_beats}. Almost all maximum lags occur within a window of $0.1$ seconds. To provide further evidence, I compute the maximum accuracy a model could attain if predicting chords at the beat level. This is done by iterating over each beat interval and assigning the chord with maximum overlap with the ground truth. This yields an accuracy of $97.1\%$. With these observations combined, I am satisfied that the estimated beats are accurate enough to be used.

To calculate features for a beat interval, I average all CQT features whose centre is contained within the interval. The CQT is calculated using a hop length of $1024$. The shorter hop length is used to minimise the effect of CQT frames with partial overlap with two beat intervals and ensures that each beat has many CQT frames associated with it. These representations have the added benefit of decreased computational cost as beats have a lower frequency than frames.


I test the model with different divisions and groupings of beats. This tests the assumption that whole beats are fine-grained enough for the model to make good predictions. I include tests where beat intervals are sub-divided into two or four, or beat intervals are joined into groups of two. I refer to this as the \emph{beat division}. I also test a \emph{perfect} beat division where the true chord transitions are taken as the beats. This is not a fair comparison as the model should not have access to true transition timings. However, it does provide an idea of how the model would fare if it could perfectly predict chord transitions. HMM smoothing is removed for beat-wise models as it is no longer necessary.

Results are shown in Table~\ref{tab:beat_division}. Using beat-wise predictions does not affect performance compared to frame-wise predictions. The model performs just as well with a beat division of $1$ as with a beat division of $1/2$ or $1/4$. The model performs worse with a beat division of $2$ though it only loses $6\%$ accuracy. This suggests that most chord transitions occur at frequencies smaller than the beats produced by \texttt{madmom} but that sometimes the two are misaligned.

The model with `perfect' beat intervals performs slightly worse on accuracy metrics but attains a very high \texttt{mirex} score of $90.4\%$,a which is the highest of any seen in the literature. This is a very promising result. Data from the CQT producing a \texttt{mirex} score of $90.4\%$ provides hope for significant improvements in the field. Why the model performs worse on accuracy metrics is not clear. It may be because averaged features from longer time periods provide better information as to the pitch classes present but dampen signal regarding the root note. Indeed, the mean chord duration is $1.68$ seconds while a CQT frame is $0.093$ seconds. Why this effect is not observed when predicting at the beat level is also not clear. Further analysis may reduce the gap between \texttt{mirex} and accuracy and improvements on ACR.

\begin{table}[H]
    \centering
    \begin{tabular}{lccccc}
        \toprule
        beat division & acc & root & third & seventh & mirex \\  
        \midrule
        1/4                 & 62.3           & 81.3          & 78.3          & 64.5           & 79.6         \\
        1/2                 & \textbf{62.8}  & 81.5          & \textbf{78.6} & \textbf{65.1}  & 79.9         \\
        1                   & 62.3           & 81.3          & 78.1          & 64.6           & 80.0         \\
        2                   & 56.4           & 74.7          & 71.6          & 58.5           & 73.4         \\
        none                & 62.5           & \textbf{81.7} & 78.2          & 64.7           & 80.2         \\
        perfect             & 61.1           & 79.6          & 76.1          & 63.4           & \textbf{90.4}\\
        \bottomrule
    \end{tabular}
    \caption{Results for different beat divisions. The beat division of `none' refers to a frame-wise \emph{CRNN} with a hop length of $4096$ and HMM smoothing applied and `perfect' refers to intervals that are calculated from the labels. Note that this `perfect' model is not a fair test. Beat-wise predictions with beat intervals of $1$ and below show performance decrease compared to frame-wise predictions. Beat intervals greater than $1$ increasingly suffer from being forced to assign predictions to larger periods of time that may align with true chord transitions. Interestingly, a beat interval of 2 still performs relatively well. A notable result is the \texttt{mirex} score of the `perfect' model. A score of $90.4\%$ is the highest of any in the literature. Despite this, its accuracy does not improve. }\label{tab:beat_division}
\end{table}

\section{Final Results}\label{sec:test-set}

For final results, I retrain select models on the combined training and validation splits and test on the held-out test split. This is an 80/20\% train/test split. I consider the original \emph{CRNN} with no improvements, the \emph{CRNN} with a weighted and structured loss and HMM smoothing, concatenating generative features with CQTs, pitch augmentation, beat-wise predictions, `perfect' beat-wise predictions and training on synthetic data. 

Results are shown in Table~\ref{tab:test_set}. Observations are largely similar to those found previously. Weighted and structured loss with smoothing improves accuracy by $1.2\%$ and pitch shifting improves accuracy by a further $1\%$. Generative features do not help and synthetic data improves performance by an additional $0.6\%$. This alone is not a clear enough signal that training on synthetic data is better, but it provides hope for further work. Finally, beat-wise predictions maintain the same performance. The `perfect' model achieves the highest performance across all metrics. The \texttt{mirex} of $90\%$ found on the validation set has reduced to $88.7\%$, and the gap with accuracy has narrowed compared to previous results in Table~\ref{tab:beat_division}. This is still a significant result and suggests that there is hope for breaking through the `glass ceiling'. The mean class-wise accuracy does not improve past $19.7\%$ without `perfect' beats. The median in this case is $6\%$. The model's performance on the long tail remains poor. Using a greater $\alpha$ in weighting may improve the picture but would require sacrificing accuracy.

Another observation is that the model's accuracy did not improve on the test set with the additional training data. Accuracy of the \emph{CRNN} with pitch shifts trained on only $60\%$ of \emph{pop} attained an accuracy of $64.3\%$, while the model trained on the full $80\%$ training split attains $63.7\%$. The discrepancy suggests that more data from the same distribution may not improve performance. 

\begin{table}[H]
    \centering
    \begin{tabular}{lcccccc}
        \toprule
         & acc & root & third & seventh & mirex & acc\textsubscript{class} \\
        \midrule
        \emph{CRNN}               & 61.6  & 79.3  & 76.5  & 63.3  & 80.6  & 18.9 \\
        + weighted/structured/HMM & 62.8  & 80.9  & 78.3  & 64.6  & 80.6  & 18.6 \\
        + gen features           & 62.7  & 80.4  & 77.9  & 64.6  & \textbf{80.6}  & 19.5 \\
        + pitch shift           & 63.8  & \textbf{82.4}  & 79.4  & 65.6  & 80.3  & \textbf{19.7} \\
        + synthetic data         & \textbf{64.4}  & 82.2  & \textbf{79.9}  & \textbf{66.3}  & \textbf{81.5}  & 18.3 \\
        + beats & 63.7  & 82.4  & 79.7  & 65.6  & 80.9  & \textbf{19.7} \\
        \midrule
        + perfect beats          & 65.8  & 84.5  & 81.7  & 67.6  & 88.7  & 21.2 \\
        \bottomrule
    \end{tabular}
    \caption{Final results from various experimental setups on the test set. The `perfect beats' model assumes oracle beat tracking and achieves the highest results across all metrics. It is excluded when considering the best results for each metric as it is not a fair comparison. Beat-wise models do not use synthetic data. Observations echo those found previously on the validation set. Adding the weighted loss, structured loss term and an HMM smoother improves accuracy by $1.2\%$. Generative features do not help. Pitch shifting improves metrics by $1\%$ and synthetic data by a further $0.6\%$.}\label{tab:test_set}
\end{table}

\section{Qualitative Analysis}

To conclude this section, I present two examples of the model's predictions from the test set using the `+ synthetic data' model. Finally, I compare beat-wise predictions with frame-wise predictions on songs not contained in the dataset.

\begin{figure}[H]
    \centering
    \includegraphics[width=1.0\textwidth]{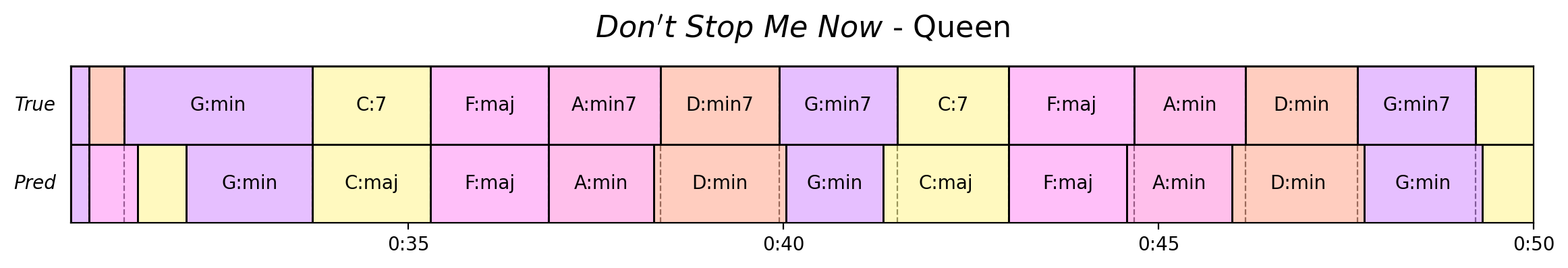}
    \caption{Comparison of predictions with ground truth on a section of \emph{Don't Stop Me Now} by Queen. Predictions have the correct root and third and are well timed but have discrepancies in predictions of sevenths. }\label{fig:dontstopmenow}
\end{figure}

The first song is \emph{Don't Stop Me Now} by Queen. With stable vocals and a clear piano part, the model fares well. The model's predictions are shown in Figure~\ref{fig:dontstopmenow}. Transitions are close to the ground truth timings. Almost all chords have the correct root and third, while sevenths are often missed. The \texttt{root} and \texttt{mirex} on this song are both $\approx 87\%$ while the overall accuracy is only $56.2\%$.

\begin{figure}[H]
    \centering
    \includegraphics[width=1.0\textwidth]{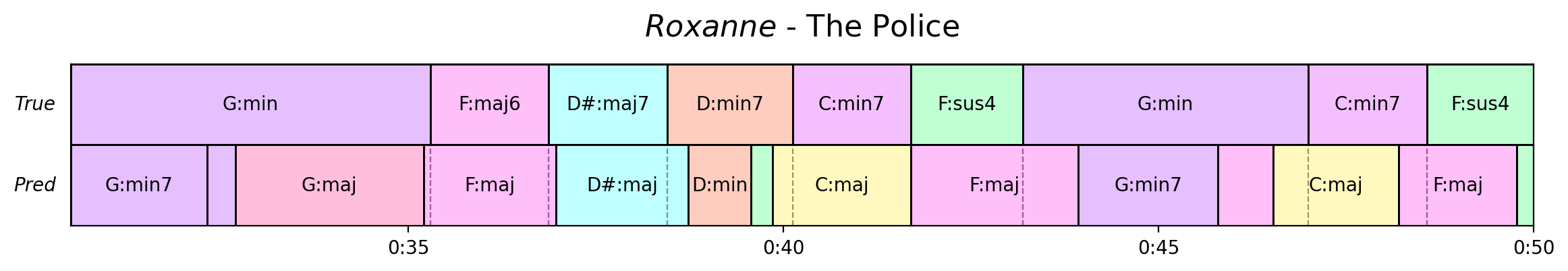}
    \caption{Comparison of predictions with ground truth on a section of \emph{Roxanne} by The Police. Predictions have the correct root but are not always well-timed and do not always have the correct quality. The model's annotation of this song is not very good. }\label{fig:roxanne}
\end{figure}

The second song is \emph{Roxanne} by The Police, illustrated in Figure~\ref{fig:roxanne}. Syncopation, ambiguous bass, and sliding vocals make this song harder to annotate. Root recall is almost $80\%$ but \texttt{mirex} is only $64\%$. Sevenths are often omitted and thirds are sometimes wrong as well. The model confuses major and minor, as well as sus4 and major qualities. There are also some predicted chord transitions that are not present in the ground truth and chord transitions are not all well-timed.

Overall, outputs are much smoother than found previously in Section~\ref{sec:crnn_examples} with more cohesive predictions of the same chord for a series of frames. However, the problem associated with identifying rarer chord qualities remains. Performance is also still highly song-dependent. The model's accuracies over songs in the test set have a standard deviation of $20\%$. 

\begin{figure}[H]
    \centering
    \includegraphics[width=1.0\textwidth]{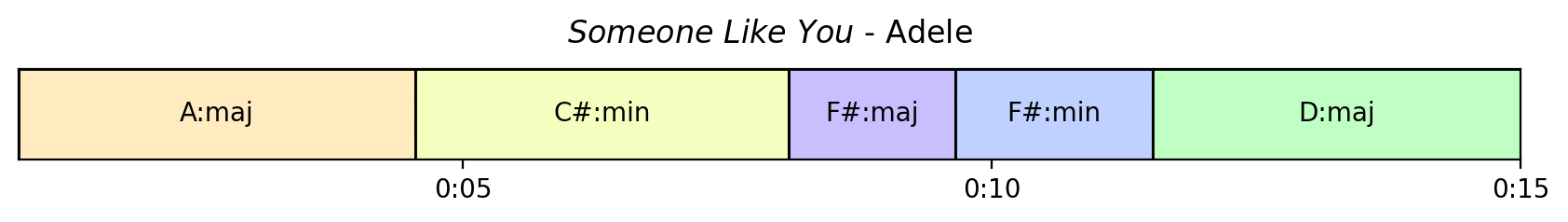}
    \includegraphics[width=1.0\textwidth]{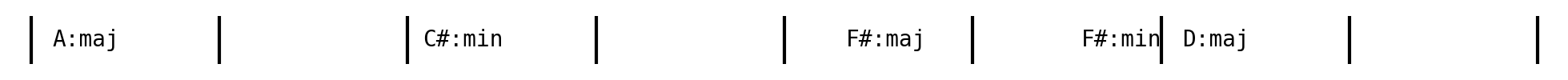}\\
    \vspace{0.2cm}
    \includegraphics[width=1.0\textwidth]{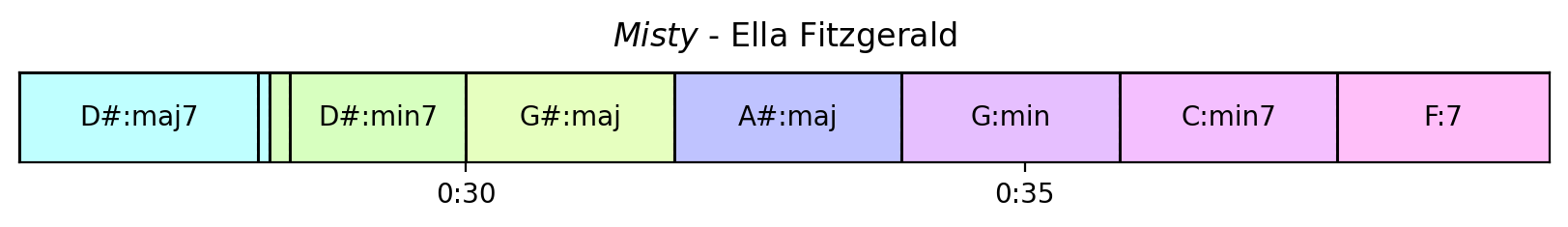}
    \includegraphics[width=1.0\textwidth]{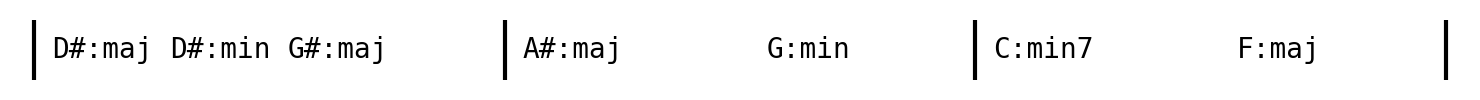}
    \caption{Comparison of frame-wise and beat-wise predictions on \emph{Someone Like You} by Adele and \emph{Misty} by Ella Fitzgerald. Beat-wise predictions are visualised over bars with a meter of 4. For both methods, adjacent predictions of the same chord are grouped together. While beat-wise predictions are more musically meaningful and easily parsed, they can be messy and hard to interpret when chord transitions are late or off the beat.}\label{fig:frame_vs_beat_wise}
\end{figure}

As a final example, frame-wise and beat-wise predictions are compared on two songs not part of the dataset in Figure~\ref{fig:frame_vs_beat_wise}: \emph{Someone Like You} by Adele and \emph{Misty} by Ella Fitzgerald. In both cases, similar chord information is visualised differently. From a musician's perspective, frame-wise predictions lead to ambiguity over how long a chord lasts in beats. Frame‑wise block lengths hint at beats but are not uniform. With time changes, this would become more problematic. In regions with rapid chord transitions, chords are not musically interpretable. In contrast, beat-wise predictions resemble musical notation more closely, and longer beat intervals prevent rapid chord changes. However, beat-wise output can still be hard to interpret when predictions occur part-way through bars.

%% file: chapters/conclusions.tex
\chapter{Conclusions and Further Work}

\section{Conclusions}

In this thesis, I have presented a thorough analysis of deep learning in automatic chord recognition. There are a few key takeaways.

ACR models are not complex. Good performance relative to state-of-the-art can be achieved with few parameters. It is likely that the task of determining which pitch classes are present from a CQT is a relatively simple operation for a neural network to learn. Performance does not increase with model size past a low threshold.

There are several explanations for the low ceiling on performance and the gap between \texttt{mirex} score and accuracy. First, annotations are too ambiguous or inconsistent for classifiers to learn the upper extensions of chords. Further research on inter-annotator agreement on this dataset is required to assess whether or not this is the case. Second, there are too few instances of rare chord classes. This leads to the current models failing to learn signals indicating the presence of such classes. Third, the current models are unable to use information from a wider context to discern chord qualities. Different genres or repetitions of the same chord within a song may give further clues about the musically coherent chord quality. Whatever the reason, chord recognition models are unlikely to become more useful than crowd-sourced annotations without addressing this issue. 

Without smoothing, frame-wise predictions result in too many chord transitions. Of the smoothing methods tested, fixed transition matrices are preferred. Weighting the loss function allows control over performance on rare qualities but requires sacrificing overall accuracy. Introducing structured representations of chords as additional targets provides a small performance gain. Features extracted from MusicGen contain information relevant to ACR but not any more than is already contained in the CQT.

Pitch augmentation works well to encourage root-invariance and improve accuracy. The use of synthetic data provides an exciting avenue for future research. Results presented here show signs that with newer models and more careful construction, synthetic data could provide many new training examples with a customisable chord distribution.

Predicting chords over beats instead of frames improves the interpretation of the model's outputs while performance is unaffected. Predicting chords over the true chord intervals results in the highest \texttt{mirex} score seen in the literature, suggesting that there are gains to be had through accurately detecting chord transitions.

While deep learning models are powerful chord recognisers, much work remains before the problem is solved. The `glass ceiling' has yet to be broken but the work presented here provides a solid foundation for future research and hope that the true ceiling is much higher.

\section{Future Work}

Many of the experiments conducted would benefit from further analysis. Implementing a sampling method which prioritises rare qualities may yield improved results over a weighted loss function. Looking at alternative methods of structuring chords beyond the pitch classes present may improve results, like the work of \citet{ACRLargeVocab1}. Larger generative models trained on a broader variety of songs may produce better representations for ACR. The work presented here also highlights new avenues of research.

\textbf{Multiple Data Sources.} Results on synthetic data show enough promise to continue this line of research. A more closely controlled chord sequence generation process may help. For example, one could construct examples designed to teach the differences between different seventh qualities and look at the effect on recall on seventh qualities to see if they improve. Other datasets also exist such as \emph{HookTheory}. I was not able to obtain audio from this source. However, results here suggest that gathering more data from the same distribution may not help. A better data source might be \emph{JAAH}, which would enable comparisons across genres and chord distributions.

\textbf{Finding better chord transitions than beats.} The high \texttt{mirex} score found in Section~\ref{sec:beat-synchronisation} suggests two things. First, targeting the problem of identifying chord transitions rather than beats may yield better results. \citet{ChorusAlignmentJAAH} jointly estimate beats and chords, but to the best of my knowledge, no modern work has jointly estimated chord transitions and chord symbols. Second, current models are missing information regarding the presence of pitch classes that are present in the CQT. Perhaps this information is spread out in time or obscured by nearby frames that are irrelevant to the current chord. Understanding this effect may lead to new insights.

\textbf{Subjective annotations.} Inter-annotator agreement of the root of a chord is estimated at lying between 76\%~\citep{AnnotatorAgreement76} and 94\%~\citep{RockHarmonyAnalysis94} but these metrics are calculated using only four and two annotators, respectively. \citet{FourTimelyInsights} posit that agreement between annotations can be far lower than that for some songs. Analysis of such an effect on commonly used datasets would provide a valuable contribution to the field. Such analysis could be used to inform the design of more subtle chord annotations that take multiple annotations and uncertainty into account.

A statement regarding the limitations of the conclusions presented here and ethics of musical machine learning models can be found in Appendix~\ref{app:limitations_and_ethics}.

%% file: chapters/appendix.tex
\appendix

\chapter{Appendix}

\section{Limitations and Ethics Statement}\label{app:limitations_and_ethics}

It is worth bearing in mind the limitations of the work presented here and its conclusions. The dataset used focuses largely on pop and rock. Cross-genre generalisation is not considered. These models expect music in standard tuning in the Wester chromatic scale. Chords themselves are also a highly Western concept. Not all music can be well described by the harmony structures considered in the chord vocabulary in this work. 

There are also ethical issues relevant to training musical machine learning models worth mentioning. The audio data in this work is subject to copyright. Research on chord recognition falls under fair use but distribution of such data should be carefully controlled. 

The generative models also present an ethical dilemma. Such models can be trained on copyrighted data without proper legal agreements with rights holder. The authors of the MusicGen models used in this work claim to have addressed such concerns. I do not support the use of work in chord recognition as a basis for advancing music generation models without proper legal and ethical issues being addressed.

\section{Weighted Chord Symbol Recall Definitions}\label{app:weighted_chord_symbol_recall_definitions}

A formal definition of WCSR is provided in Equation~\ref{eq:wcsr}. The WCSR is a measure of the percentage of time that the model's predictions are correct.

\begin{equation}\label{eq:wcsr}
    WCSR = 100\cdot\frac{1}{Z}\sum_{i=1}^{N} \int_{t=0}^{T_i} M(y_{i,t},\hat{y}_{i,t}) dt
\end{equation}
\begin{equation}
    Z = \sum_{i=1}^{N} \int_{t=0}^{T_i} \mathbb{I}_M(y_{i,t}) dt
\end{equation}

where $M(y, \hat{y})\in\{0,1\}$ is the measure of correctness which varies across metrics. For example, $M(y, \hat{y})$ for \texttt{root} equals $1$ if $y$ and $\hat{y}$ share the same root and $0$ otherwise. $N$ is the number of songs, $T_i$ is the length of song $i$, $y_{i,t}$ is the true chord at time $t$ of song $i$, and $\hat{y}_{i,t}$ is the predicted chord at time $t$ of song $i$. $Z$ normalises by the length of time for which the metric $M$ is defined. This is necessary as \texttt{X} symbols are ignored and \texttt{seventh} ignores some qualities. Further details can be found in the \texttt{mir\_eval} documentation. $\mathbb{I}_M(y_{i,t})=1$ if $M$ is defined for label $y_{i,t}$ and $0$ otherwise. Finally, we multiply by 100 to convert to a percentage.

I also define WCSR for a single class $c$ in Equation~\ref{eq:wcsr_c}. This is useful for understanding the performance of the model on a specific chord class.

\begin{equation}\label{eq:wcsr_c}
    \text{WCSR}(c) = \frac{1}{Z_c}\sum_{i=1}^{N} \int_{t=0}^{T_i} M(y_{i,t},\hat{y}_{i,t}) \cdot \mathbb{I}_c(y_{i,t}) dt
\end{equation}
\begin{equation}
    Z_c = \sum_{i=1}^{N} \int_{t=0}^{T_i} \mathbb{I}_M(y_{i,t})\cdot \mathbb{I}_c(y_{i,t}) dt
\end{equation}

where $N$, $T$, $M$, $y_{i,t}$, $\hat{y}_{i,t}$ and $\mathbb{I}_M(y_{i,t})$ are defined as before in Equation~\ref{eq:wcsr}. $\mathbb{I}_c(y_{i,t})$ is $1$ if the true chord at time $t$ of song $i$ is class $c$ abd $0$ otherwise. $Z_c$ normalises by the length of time for which the chord $c$ is playing and for which the metric $M$ is defined, in a similar fashion to $Z$ in Equation~\ref{eq:wcsr}.

\section{Cross Correlation for Alignment Verification}\label{app:cross_correlation}

\begin{figure}[H]
    \centering
    \includegraphics[width=1.0\textwidth]{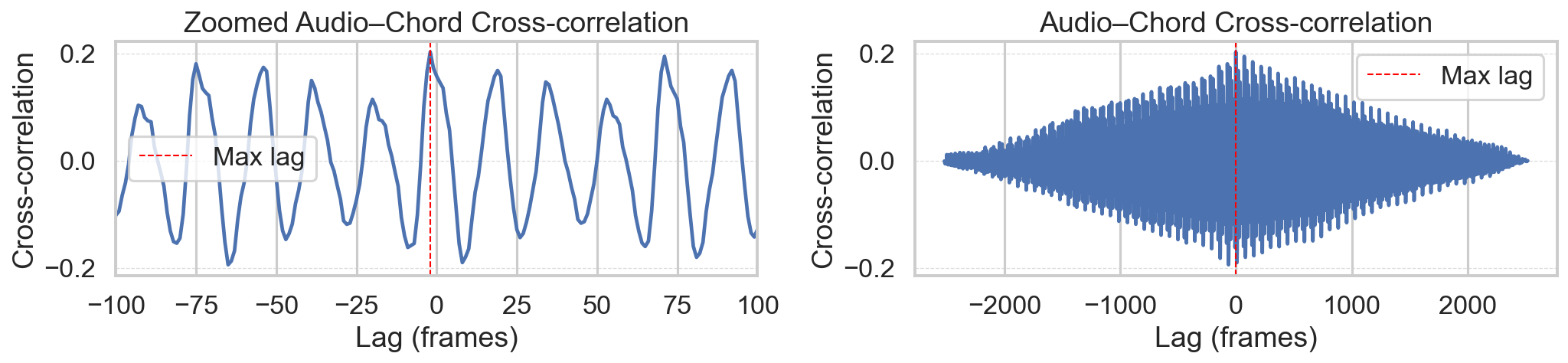}
    \caption{Cross-correlation of the derivative of the CQT of the audio and the chord annotations for a single song. We can see correlation peaking in regular intervals of around 20 frames. 1 frame is $93$ms so 20 frames $\approx 1.86$ seconds. Zooming out, we observe peaks in correlation centred around 0.}\label{fig:cross-correlation}
\end{figure}

\section{Learning Rate and Scheduler Experiment Results}\label{app:learning_experiment_results}

\begin{figure}[H]
    \centering
    \includegraphics[width=1.0\textwidth]{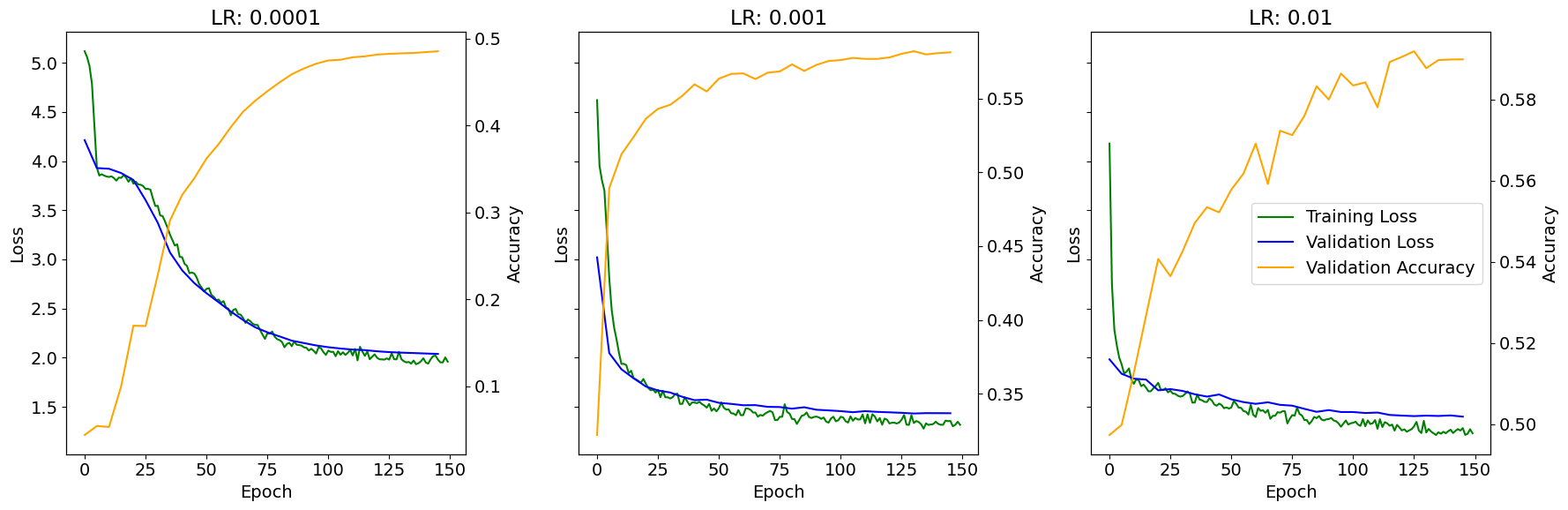}
    \caption{Training graphs for the \emph{CRNN} with different learning rates and the \texttt{cosine} scheduler. The learning rate of $0.001$ seems to be the best, as it converges in a reasonable time and the validation accuracy increases in a stable fashion. While it may seem that running for more epochs may increase performance, this was not found to be the case empirically. The best model was often achieved around epoch 100.}\label{fig:lr_search_cosine}
\end{figure}

\begin{table}[H]
    \centering
    \begin{tabular}{lcccccc}
        \toprule
        lr & scheduler & acc & root & third & seventh & mirex \\
        \midrule
        0.01 & Cosine &  53.6 & 69.5 & 66.9 & 55.7 & 78.6 \\
        0.001 & Cosine & \emph{59.7} & \emph{78.3} & \emph{75.0} & \emph{62.0} & \emph{\textbf{79.8}} \\
        0.0001 & Cosine & 53.2 & 72.1 & 66.9 & 55.2 & 72.0 \\
        \midrule
        0.001 & Plateau & \textbf{59.9} & 78.4 & 75.2 & \textbf{62.2} & 79.7 \\
        0.001 & None & 59.8 & \textbf{78.7} &\textbf{75.5} & 62.0 & 78.8 \\
        \bottomrule
    \end{tabular}
    \caption{\emph{CRNN} model results with different learning rates and schedulers. Best results over learning rates are \emph{italicised} and best results over schedulers are in \textbf{boldface}. A learning rate of $0.001$ performs the best on all metrics. The differences between learning rate schedulers are so small that the choice between them is arbitrary. }\label{tab:crnn_lr}
\end{table}

\section{Revisiting the Spectrogram}\label{sec:spectrogram-results}

\subsection{Spectrogram Variants}\label{sec:spectrogram-variants}

It is standard practice in ACR to use a CQT as input. However, \citet{20YearsofACR} raise the question of whether the CQT is truly the best choice. They suggest that the pitch-folding of the CQT may distort the harmonic structure of notes. By contrast, \citet{MelodyTranscriptionViaGenerativePreTraining} use a mel-spectrogram in place of a CQT. 

I test four spectrogram variants in Table~\ref{tab:spectrograms}. These include the standard CQT, mel-spectrogram and linear spectrogram. I also calculate a chroma-CQT to test whether the model is using information from multiple octaves better than a hand-crafted algorithm. The chroma-CQT is calculated by summing CQT values across octaves. Spectrogram calculations are all implemented in \texttt{librosa}~\citep{librosa}. I use $216$ bins for the CQT and mel spectrograms and $2048$ fast Fourier transform (FFT) bins for the linear spectrogram with a hop length of $4096$ for all. 

Results show that CQTs are the best choice. This raises questions as to the validity of the conclusions drawn by \citet{MelodyTranscriptionViaGenerativePreTraining}. They claim that their generative features are better than hand-crafted features. However, they only compare to mel-spectrograms which may not perform as well as CQTs for the related task of melody recognition. The CQT is also better the chroma-CQT. We can be confident that the model is using information from multiple octaves more efficiently than the simply summing across octaves.

\begin{table}[h]
    \centering
    \begin{tabular}{lccccc}
        \toprule
        spectrogram & acc & root & third & seventh & mirex \\  
        \midrule
        CQT & \textbf{60.2} & \textbf{78.4} & \textbf{75.3} & \textbf{62.5} & \textbf{79.5} \\
        chroma-CQT & 50.1 & 71.4 & 65.7 & 52.0 & 69.8 \\
        mel & 52.7 & 69.1 & 66.3 & 54.6 & 70.6 \\
        linear & 51.2 & 66.1 & 63.0 & 53.1 & 73.8 \\
        \bottomrule
    \end{tabular}
    \caption{Results for CQT, chroma-CQT, mel and linear spectrograms. The CQT is certainly the best feature. The other all perform similarly on accuracy and \text{mirex}, but the chroma-CQT does comparatively at identifying thirds and sevenths. }\label{tab:spectrograms}
\end{table}

\subsection{Hop Lengths}\label{sec:hop-lengths}

Different hop lengths have been used to calculate the CQT ranging from 512~\cite{ACRLargeVocab1} up to 4096~\citep{StructuredTraining}. In previous experiments I have used a hop length of $4096$ as is used by the authors of \emph{CRNN}~\citep{StructuredTraining}. Shorter frames would reduce the number of transition frames but require more computational cost. If frame lengths are too short, the Fourier transform may not be able to capture the harmonic structure of the audio.

In Table~\ref{tab:hop_lengths}, I test the effect of different hop lengths on the model's performance. I use a CQT with $216$ bins and a hop length of $512$, $1024$, $2048$, $4096$, $8192$ and $16384$. Results indicate that performance is similar for hop lengths of $4096$ and below. Performance suffers for greater hop lengths. While it could be argued that $2048$ does better than $4096$, this difference is small enough that it is not worth the increased computational cost. Models trained with a hop size of $2048$ take at least twice as long to train and evaluate as those trained on a hop size of $4096$.

\begin{table}[h]
    \centering
    \begin{tabular}{lccccc}
        \toprule
        hop length & acc & root & third & seventh & mirex \\  
        \midrule
        512 & 60.1 & 78.3 & \textbf{75.5} & 62.4 & \textbf{80.0} \\
        1024 & 60.2 & \textbf{78.7} & 75.2 & 62.5 & 79.6 \\
        2048 & \textbf{60.3} & 78.5 & 75.2 & \textbf{62.6} & 79.6 \\
        4096 & 60.0 & 78.1 & 75.0 & 62.2 & 79.2 \\
        8192 & 57.9 & 76.2 & 72.9 & 60.1 & 79.3 \\
        16384 & 53.3 & 71.7 & 68.0 & 55.4 & 77.9 \\
        \bottomrule
    \end{tabular}
    \caption{Results over different hop lengths for CQT calculation. Any hop length in the range $512$ to $4096$ has similar performance. For frames that are any longer, performance suffers. This is likely caused by the requirement for the model to assign a single chord class to each frame. The longer the frame, the greater potential there is for multiple chords to be playing during the frame. }\label{tab:hop_lengths}
\end{table}

\section{Small vs Large Vocabulary}\label{app:small_vs_large_vocabulary}

Some initial experiments were conducted over a smaller vocabulary with $C=26$. This vocabulary includes a symbol for major and minor over each root and two special symbols, \texttt{N} and \texttt{X} for no chord and unknown chord respectively. This contrasts the much larger vocabulary with 14 chord qualities for each root which is used for the majority of the experiments. With this larger vocabulary, $C=170$.

Table~\ref{tab:small_vs_large_vocab} shows the results of the \emph{CRNN} trained over the smaller and larger vocabulary evaluated on the small vocabulary. The predictions of the model and the reference labels are all mapped to the small vocabulary before being evaluated in the same was as described in Section~\ref{sec:evaluation}. This test was to verify that the model trained on the larger vocabulary performs well competitively on the smaller vocabulary. If the model trained on the larger vocabulary performed poorly on the smaller vocabulary, it may be prudent to first try to improve performance on this smaller vocabulary. It may also be a sign that the larger vocabulary is too complex or that the more detailed annotations are inconsistent.

However, the table shows very similar performance between both models. This allows us to proceed with the larger vocabulary for the rest of the experiments. The larger vocabulary is also more consistent with the literature and allows for a model to produce far more interesting chord predictions than simply minor, major and root. 

\begin{table}[H]
    \centering
    \begin{tabular}{lccc}
        \toprule
        Vocab & $C$ & acc & root \\  
        \midrule
        small & $26$ & 76.7 & 80.1 \\
        large & $170$ & 76.0 & 79.1 \\
        \bottomrule
    \end{tabular}
    \caption{\emph{CRNN} with a small and large vocabulary. Metrics show similar performance between the two. Training on the large vocabulary does not prevent the model from learning how to classify the smaller vocabulary. Thus, I proceed with the larger vocabulary.}\label{tab:small_vs_large_vocab}
\end{table}

Note that the \texttt{mir\_eval} package also includes a \texttt{majmin} evaluation metric that compares chords over just the major and minor qualities. However, this is not quite the same as the test above due to subtleties in how \texttt{mir\_eval} chooses whether or not a chord is major or minor. It ends up ignoring many chords that could be mapped to these qualities in the smaller vocabulary. Coincidentally, the \emph{CRNN} with the default parameters attains a \texttt{majmin} accuracy of $76.0\%$ over the larger vocabulary. This further confirms that we need not continue to test on the smaller vocabulary. The  \texttt{majmin} metric is not used in the rest of the thesis as it is not as informative as the other metrics and the \texttt{third} metric is highly correlated with it.

\section{Chord Mapping}\label{app:chord_mapping}

Chords in Harte notation were mapped to the vocabulary with $C=170$ by first converting them to a tuple of integers using the Harte library. These integers represent pitch classes and are in the range 0 to 11 inclusive. They are transposed such that 0 is the root pitch. These pitch classes were then matched to the pitch classes of a quality in the vocabulary, similar to the work by \citet{StructuredTraining}. However, for some chords, this was not sufficient. For example, a \texttt{C:maj6(9)} chord would not fit perfectly with any of these templates due to the added 9th. Therefore, the chord was also passed through Music21's~\citep{music21} chord quality function which matches chords such as the one above to major. This function would not work alone as its list of qualities is not as rich as the one defined above. If the chord was still not matched, it was mapped to \texttt{X}. This additional step is not done by \citet{StructuredTraining} but gives more meaningful labels to roughly one third of the chords previously mapped to \texttt{X}.

\section{CRNN with CR2}\label{app:crnn_with_cr2}

\begin{table}[H]
    \centering
    \begin{tabular}{lccccccc}
        \toprule
        cr2 & acc & root & third & seventh & mirex & acc\textsubscript{class} & median\textsubscript{class} \\  
        \midrule
        on & 59.7 & \textbf{78.9} & \textbf{75.6} & 61.9 & \textbf{80.5} & 18.4 & 0.4 \\
        off & \textbf{60.2} & 78.4 & 75.3 & \textbf{62.5} & 79.5 & \textbf{19.4} & \textbf{1.1} \\
        \bottomrule
    \end{tabular}
    \caption{\emph{CRNN} with and without the added `CR2' decoder. Performance is very similar between the two. It could be argued that te model with CR2 on is better, but for simplicity, I proceed with the model without CR2. One could also argue that the effect of CR2 is similar to simply adding more layers to the GRU already present in the \emph{CRNN}.}
\end{table}

\section{A Long Run with SGD}\label{app:long_sgd}
\begin{figure}[H]
    \centering
    \includegraphics[width=1.0\textwidth]{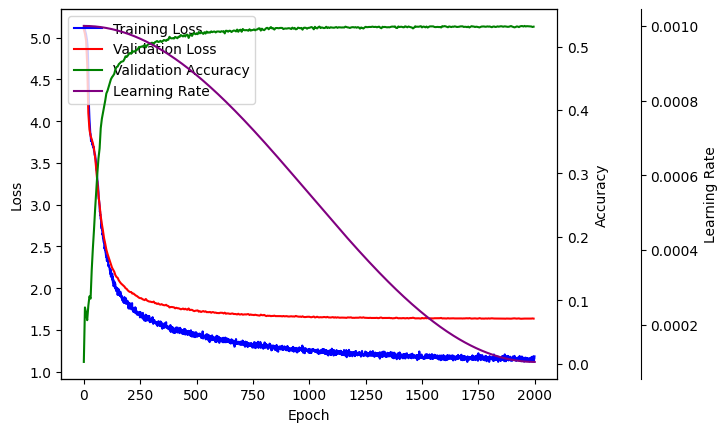}
    \caption{Training graphs for the \emph{CRNN} trained with SGD, momentum 0.9, a learning rate of $0.001$ and the \texttt{cosine} scheduling for 2000 epochs. Convergence is reached but performance does not exceed that which is achieved by Adam over 150\,epochs. Furthermore, there is signifcant computational cost associated with running for 2000 epochs. I proceed with Adam for the remainder of experiments. }\label{fig:long_sgd}
\end{figure}

\section{Random Hyperparameter Search Sets}\label{app:random_hyperparameter_search_sets}

The random hyperparameter search for the \emph{CRNN} was done over the following variables and values:
\begin{itemize}
    \item \texttt{hidden\_size} $\in$ \{32, 64, 128, 256, 512\}
    \item \texttt{num\_layers} $\in$ \{1, 2, 3\}
    \item \texttt{segment\_length} $\in$ \{5, 10, 15, 20, 25, 30, 35, 40, 45\}
    \item \texttt{kernel\_size} $\in$ \{5, 6, 7, 8, 9, 10, 11, 12, 13, 14, 15\}
    \item \texttt{cnn\_layers} $\in$ \{1,2,\ldots,5\}
    \item \texttt{cnn\_channels} $\in$ \{1,2,\ldots,5\}
\end{itemize}

For each run, a value was selected for each hyperparameter, with each possible value equally likely.



\section{Confusion Matrix of CRNN over Roots}\label{app:cm_roots}

\begin{figure}[H]
    \centering
    \includegraphics[width=1.0\textwidth]{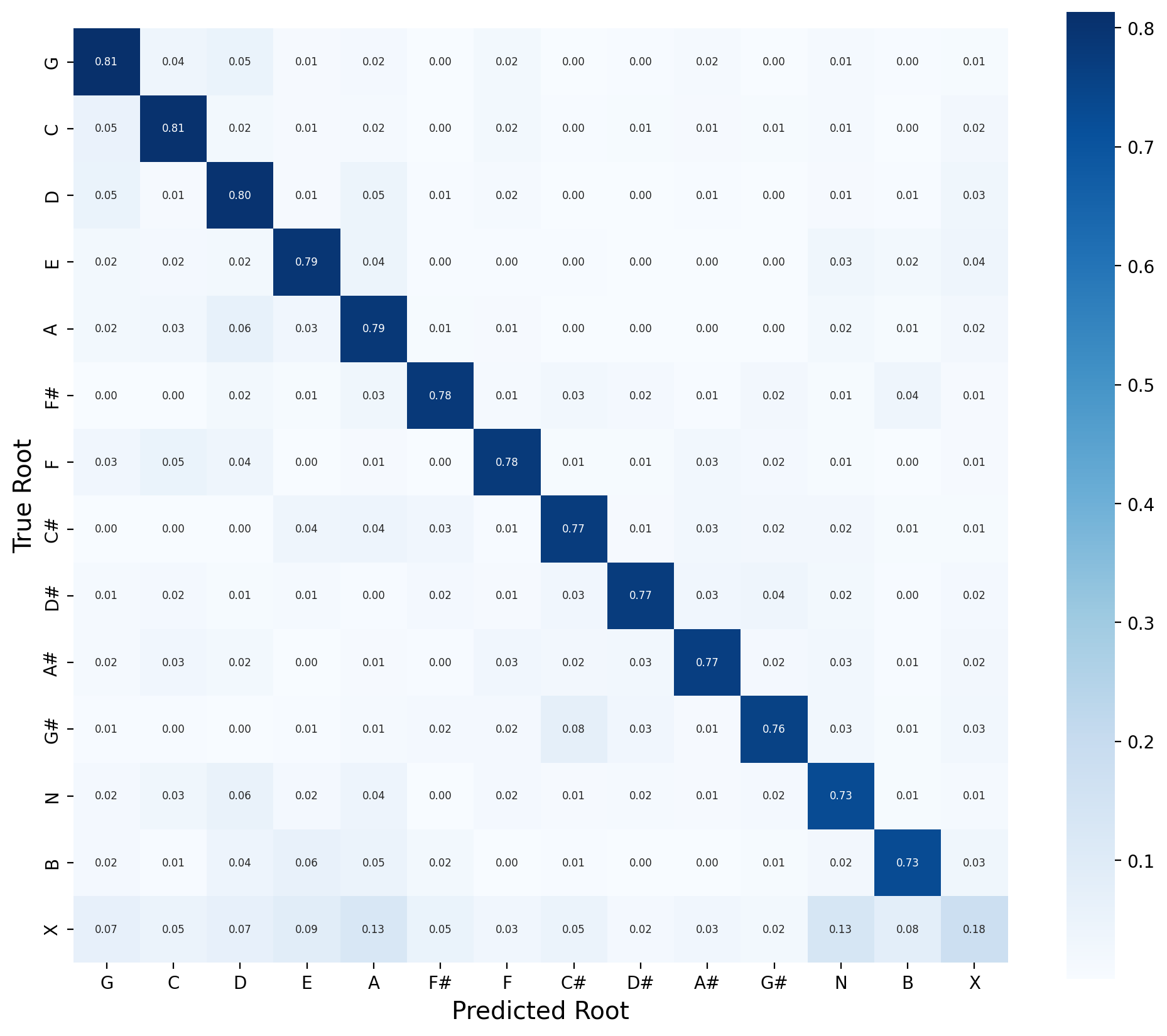}
    \caption{Performance is relatively stable across roots. The only outlier is the unknown chord symbol \texttt{X}. This is to bexpected given the ambiguous nature of the chord. }
    \label{fig:cm_roots}
\end{figure}



\section{Incorrect Region Lengths With/Without Smoothing}\label{app:histogram_over_region_lengths}

\begin{figure}[H]
    \centering
    \includegraphics[width=0.6\textwidth]{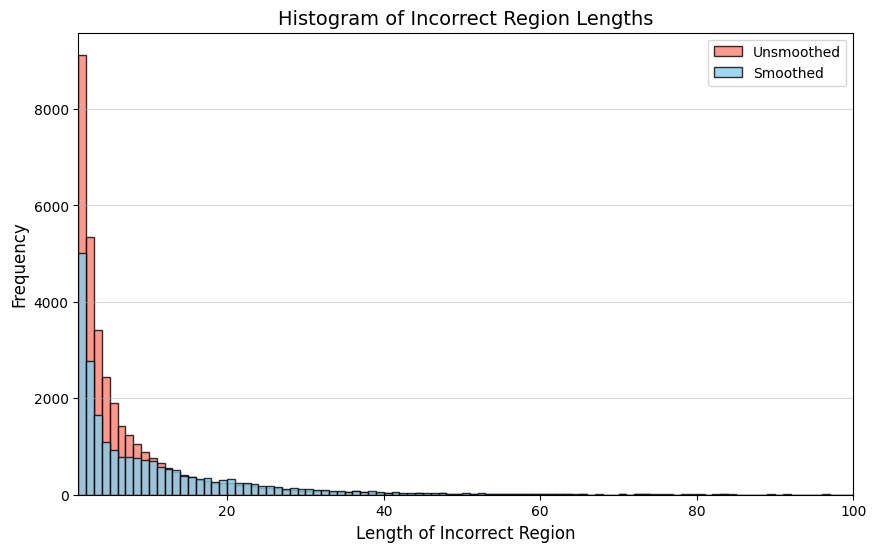}
    \caption{Histogram over incorrect region lengths for the \emph{CRNN} with and without smoothing. An incorrect region is defined as a sequence if incorrect frames with correct adjacent of either end. Both distributions have a long-tail, with $26.7\%$ regions being of length 1 without smoothing. This raises concerns over the smoothness of outputs and requires some form of post-processing explored in Section~\ref{sec:decoding}. The distribution is more uniform with smoothing, with approximately half the very short incorrect regions.}
    \label{fig:histogram_over_region_lengths}
\end{figure}

\section{Accuracy over the Context}\label{app:accuracy_over_context}
\begin{figure}[H]
    \centering
    \includegraphics[width=0.8\textwidth]{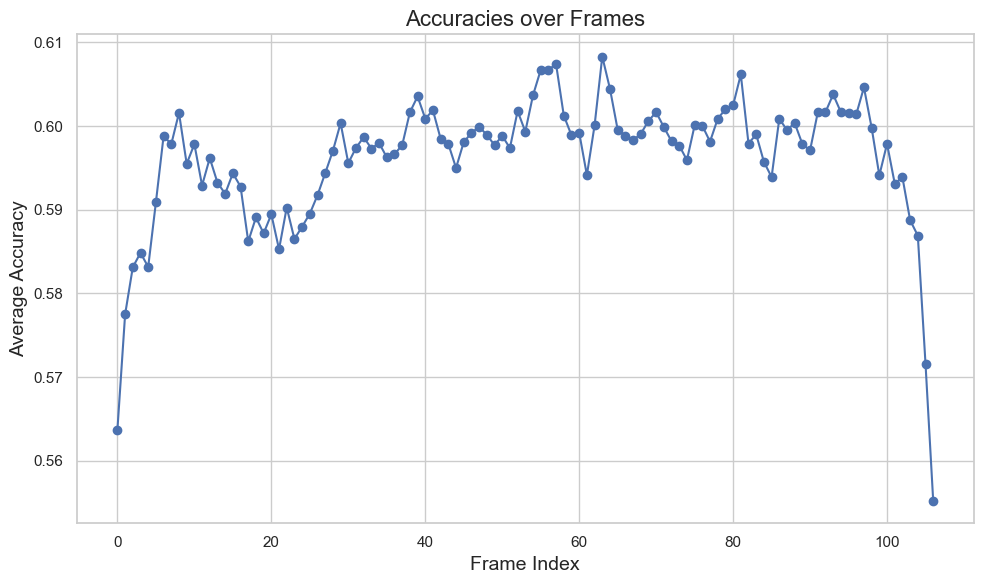}
    \caption{Average frame-wise accuracy of the \emph{CRNN} model over the patch of audio. The model performs worse at the beginning and end of the patch of audio, as expected. However, the differences are only $~0.05$. We propose that the context on one side is enough for the model to attain the vast majority of the performance attained with bi-directional context. This plot supports our procedure of evaluating over the entire song at once. }\label{fig:crnn_context}
\end{figure}

\section{Accuracy vs Context Length of Evaluation}\label{app:accuracy_vs_context_length}

\begin{figure}[H]
    \centering
    \includegraphics[width=0.5\textwidth]{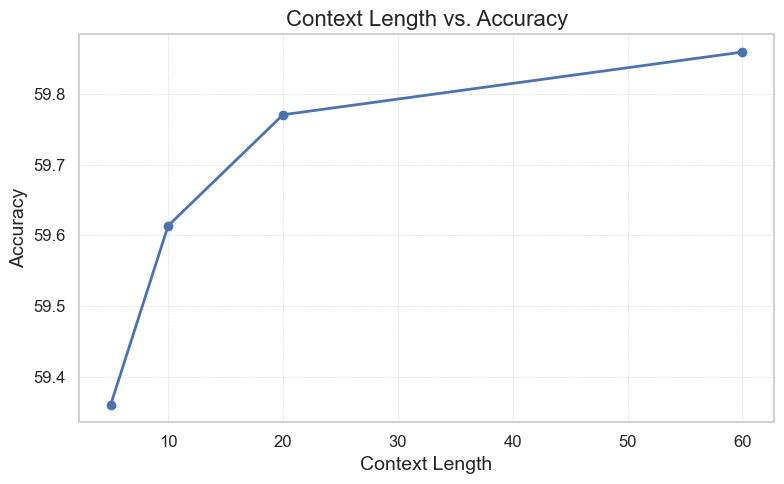}
    \caption{Accuracy with increasing segment length of validation set. The accuracy increases very slightly. I choose to continue evaluating over the entire song at once.}
    \label{fig:accuracy_vs_context_length}
\end{figure}

\section{HMM Smoothing Effect}\label{app:hmm_smoothing_effect}

\begin{figure}[H]
    \centering
    \includegraphics[width=0.9\textwidth]{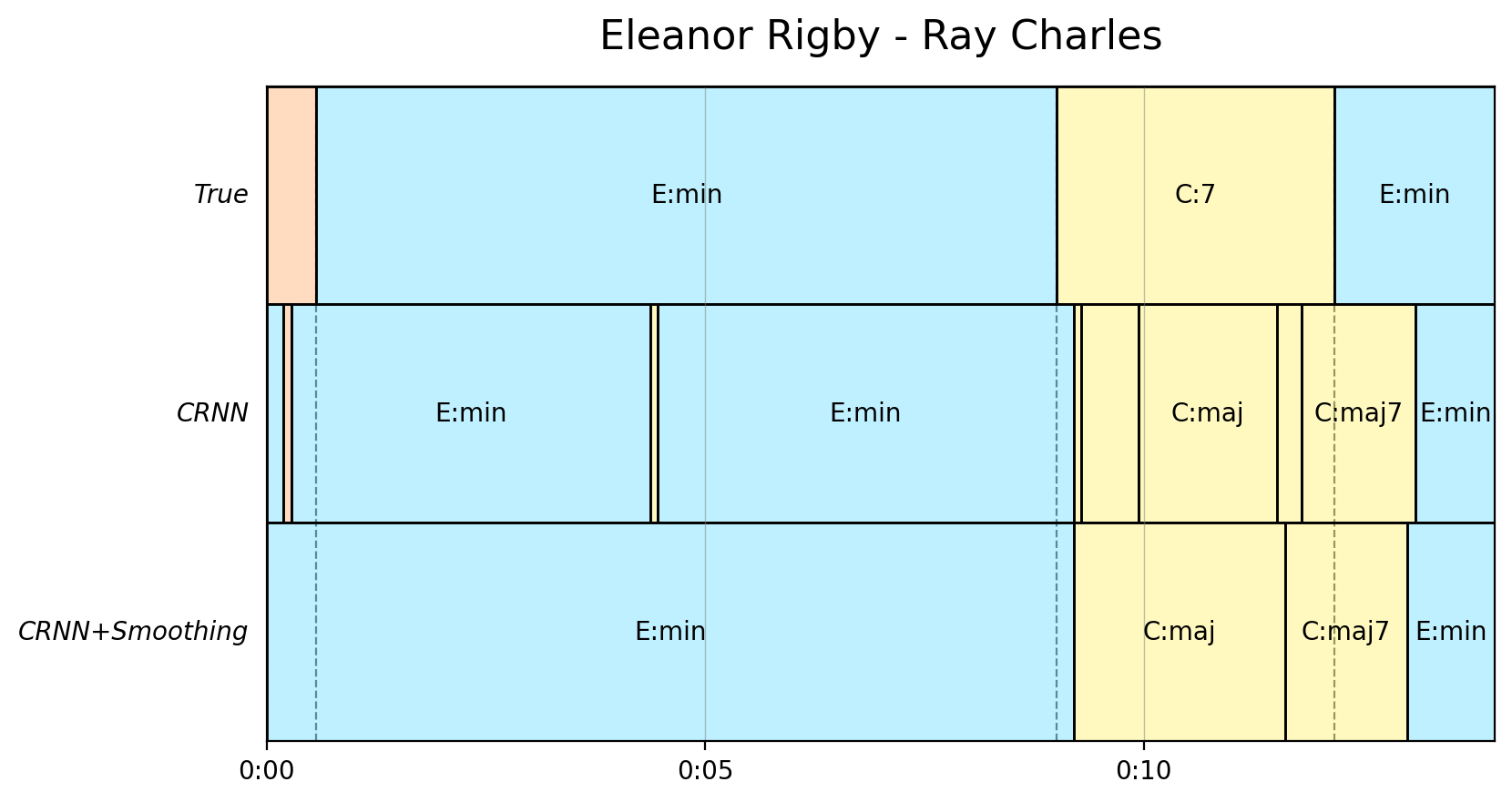}
    \caption{An example of the effect of the HMM on the predictions of the \emph{CRNN} model. The top plot shows the ground truth. The middle plot shows frame-wise predictions of the \emph{CRNN} without smoothing. The bottom plot shows the predictions after smoothing. Chords are coloured by their equivalent chord in the small vocabulary as it makes the plot easier to interpret. The original predictions contain many unnecessary and nonsensical chord transitions. These have been smoothed out by the HMM. The resulting chords appear more similar to the ground truth even if frame-wise accuracy has not changed much.}\label{fig:hmm_smoothing_example}
\end{figure}

\section{Weighted Loss Confusion Matrix}\label{app:weighted_loss_confusion_matrix}

\begin{figure}[H]
    \centering
    \hspace{-1.5cm}
    \includegraphics[width=0.9\textwidth]{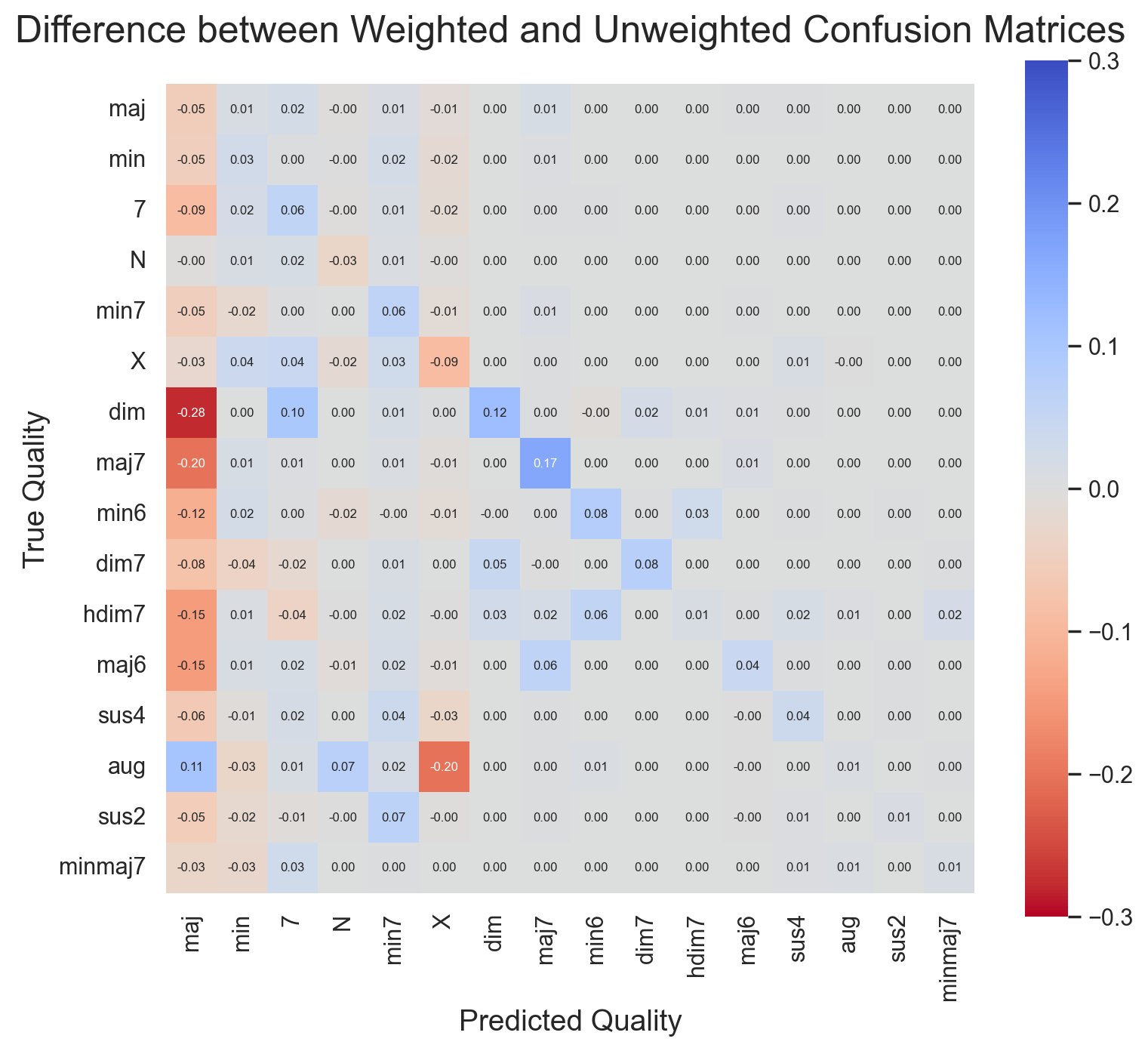}
    \caption{ Most of the diagonal entries in the confusion matrix increase. Recall on major7 qualities increases by $0.17$. The only qualities to decrease in recall are major, \texttt{N} and \texttt{X}. I conclude that weighting the loss does improve the model. The weighted model predicts \texttt{X} $2.2$ times less often. This may be how the weighted model improves class-wise metrics without sacrificing too much overall accuracy since \texttt{X} frames are ignored for evaluation.}\label{fig:hmm_smoothing_example}
\end{figure}

\section{Structured Loss Experiment Results}\label{app:structured_loss_experiment_results}
\begin{figure}[H]
    \centering
    \includegraphics[width=1.0\textwidth]{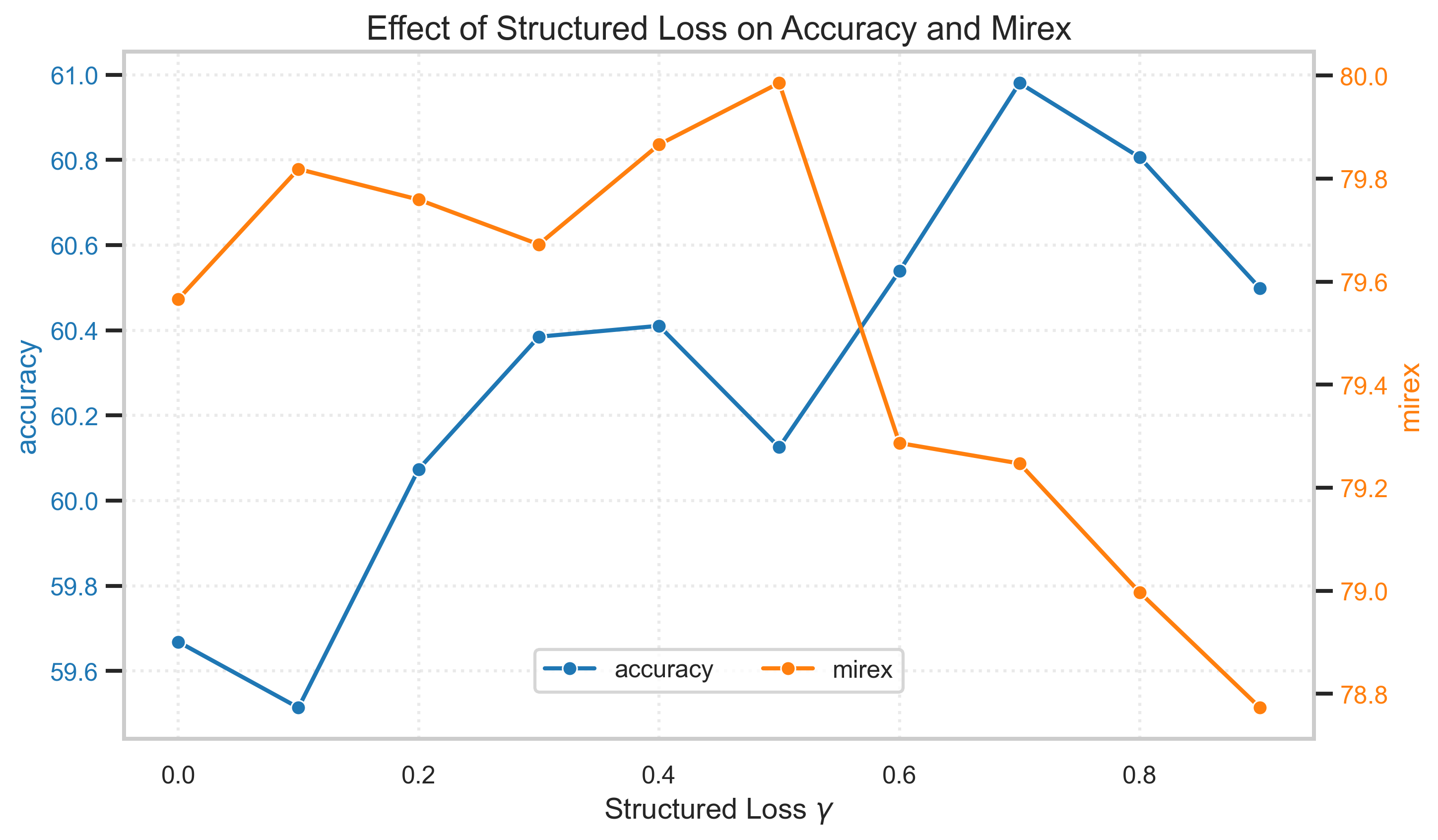}
    \caption{Effect of structured loss on the \emph{CRNN} model with varying $\gamma$. As we increase $\gamma$, accuracy improves but \texttt{mirex} behaves erratically, worsening at the higher end. The other metrics behave similarly to accuracy. I choose $\gamma=0.7$ based on peak accuracy. }\label{fig:structured_loss}
\end{figure}

\section{Details of Generative Feature Extraction}\label{app:generative_feature_extraction}

Overlapping $5$ second chunks of audio were fed through MusicGen in a batched fashion. This first requires passing the audio through the pre-trained Encodec audio tokeniser~\citep{Encodec}. These are then fed through the language model. I take the output logits as the representation for each frame. The model outputs logits in four `codebooks', each $2048$-dimensional vectors, intended to represent different granularities of detail in the audio. Audio segments are overlapped such that every frame has context from both directions. The multiple representations for each frame are averaged. Finally, these representations are upsampled. The model operates at a frame rate of 50Hz. To compute a representation with the same frame length as the CQT, I take the mean over the frames outputted by the model closest to the centre of the CQT frame. In case averaging over frames dampened the signal, I also tried linearly interpolating between the two closest frames outputted by the model. However, this was empirically found to perform slightly worse. Results are left to Appendix~\ref{app:linear_interpolation_vs_area_averaging}. This feature extraction required the use of NVIDIA RTX A6000 GPUs. The extraction process takes 4 hours for each model over the entire dataset.

\section{Upsampling Methods in Generative Feature Extraction}\label{app:linear_interpolation_vs_area_averaging}

\begin{table}[H]
    \centering
    \begin{tabular}{cccccc}
        \toprule
        upsample & accuracy & root & third & seventh & mirex  \\  
        \midrule
        area & \textbf{59.4} & \textbf{77.8} & \textbf{74.6} & \textbf{61.7} & \textbf{78.1} \\
        lerp & 58.4 & 77.7 & 74.0 & 60.6 & 78.2 \\
        \bottomrule
    \end{tabular}
    \caption{Comparison of generative feature extraction with linear interpolation and area averaging for the musicgen-large and a linear projection down to $64$ dimensions and averaging over the four codebooks. The results are very similar, but the area averaging method is slightly better in all metrics. I therefore choose to continue averaging over model frames in order to upsample to CQT frames.}\label{tab:linear_interpolation_vs_area_averaging}
\end{table}

\section{Generative Feature Low-Dimensional Projection}\label{app:projection_dimensionality}

\begin{table}[H]
    \centering
    \begin{tabular}{cccccc}
        \toprule
        $d$ & accuracy & root & third & seventh & mirex  \\  
        \midrule
        16   & 58.4       & 77.6       & \textbf{74.5} & 60.6       & 77.2       \\
        32   & 57.8       & 77.5       & 73.6          & 60.0       & 77.8       \\
        64   & \textbf{59.2} & 77.5    & 74.3          & \textbf{61.4} & \textbf{77.9} \\
        128  & 58.1       & 77.0       & 73.6          & 60.3       & 77.4       \\
        256  & 58.7       & 77.6       & 74.3          & 60.9       & 77.5       \\
        512  & 58.3       & \textbf{77.9} & 74.3      & 60.6       & 77.6       \\
        1024 & 58.4       & 76.7       & 73.5          & 60.7       & 76.7       \\ 
        \bottomrule
    \end{tabular}
    \caption{Generative feature extraction with different projection dimensions, $d$. All results use the musicgen-large model and average across the four codebooks. There are no large differences between the different dimension reductions. I take $d=64$ as it performs the best, but there is no evidence that this is not due to randomness in the optimisation process. }\label{tab:projection_dimensionality}
\end{table}

\section{Generative Features with Different Models}\label{app:generative_feature_extraction_models}

\begin{table}[H]
    \centering
    \begin{tabular}{lc}
        \toprule
        model  & acc  \\
        \midrule
        large  & \textbf{60.2} \\
        small  & 59.8 \\
        melody & 59.9 \\
        chord  & 59.7 \\
        \bottomrule
    \end{tabular}
    \caption{Results for generative features extracted from MusicGen~\citep{MusicGen} and MusiConGen~\citep{MusiConGen} models with different musicgen-models. Only accuracy is reported as the other metrics are qually similar. These models are musicgen-large, musicgen-small, musicgen-melody and MusiConGen, referred to as large, small, melody and chord respectively. There are very few differences between the models. This suggests that the small model has just as much information in its internal representations that is useful for identifying chords as the other models. Another point of note is the comparison between chord and melody. The chord model is a chord-conditioned fine-tuned version of melody. It is surprising that the chord-conditioning did not help compared to the non chord-conditioned model.}\label{tab:musicgen_pivot}
\end{table}

\section{Generative Features with Different Codebook Reductions}\label{app:generative_feature_extraction_reductions}

\begin{table}[H]
    \centering
    \begin{tabular}{lc}
        \toprule
        reduction   & accuracy \\  
        \midrule
        concat       & 58.9     \\
        codebook\_2  & 58.9     \\
        codebook\_3  & 57.5     \\
        codebook\_1  & 59.0     \\
        codebook\_0  & 58.1     \\
        avg          & \textbf{59.4}     \\
        \bottomrule
    \end{tabular}
    \caption{Accuracy for different codebook reduction methods. Other metrics are omitted as they do not provide more information. All results are for musicgen-large with reduction down to $64$ dimensions. Reductions of the form `codebook\_$n$' refer to training on codebook of index $n$ from the model. Performance is similar across reductions except for codebook\_0 and codebook\_3 which perform worse. I choose the averaging reduction based on maximum accuracy. }\label{tab:reduction_accuracy}
\end{table}

\section{Jazz Chord Progression Generation}\label{app:jazz_chord_progression_generation}

The theory of functional harmony is a set of rules that govern the relationships between chords in a piece of music. While these rules are not always followed, many chord progressions can be parsed and broken down into the rules that have been followed to create them. The rules are based on the relationships between the chords and the keys they are in. For example, a chord progression that moves from a tonic chord to a dominant chord is said to be following the rule of \emph{dominant function}. This is a common rule in jazz music and is often used to create tension and resolution in a piece of music.

I first decide whether we are in major or minor, each with probability $0.5$. I then uniformly sample a tonic from the set of notes in the Western chromatic scale. From this tonic, seven functional chords are decided before sequence generation. These are all probabilistic. For example, the tonic chord is always the tonic, but the dominant chord can be of \texttt{maj}, \texttt{7}, \texttt{sus4}, \texttt{aug} or \texttt{dim7} qualities. The probabilities are user-tuned but do not matter very much to the funcionality of the synthetic dataset. 

For chord sequence generation, various rules are followed in a probabilistic manner. Progressions have a random length, uniformly sampled in the range $[4, 10]$.

\begin{itemize}
    \item Tonic (I) may move to predominant chords (ii, IV, vi) or occasionally mediant (iii).
    \item Predominant chords (ii, IV) resolve to the dominant (V).
    \item Dominant (V) usually cadences back to tonic (I) or sometimes moves to vi.
    \item Tonic substitute (vi) leads to ii or iii.
    \item Mediants (iii) feed into vi.
    \item Unspecified or fallback transitions are routed toward ii to maintain forward motion.
\end{itemize}

A time-aligned chord sequence is then calculated in a similar format to that provided by the \texttt{jams} package. This assumes that each chord is played for one bar, that the BPM is always followed, and that MusiConGen simply loops over the chord progression if the end is reached. These assumptions were found to hold on manually inspected examples.

For further details and exact probabilities used, please refer to the provided code.\footnote{\url{https://github.com/PierreRL/LeadSheetTranscription/blob/main/src/data/synthetic_data/chord_sequence.py}}

\section{Calibration to Handle Distribution Shift}\label{app:calibration}

To correct for the distribution shift between synthetic training data and the \emph{pop} train split, I estimate the empirical class probabilities in each domain---\(P_{\text{train}}(y)\) and \(P_{\text{pop}}(y)\)---and rescale the model’s logits by the ratio
\begin{equation}
r(y)\;=\;\frac{P_{\text{pop}}(y)}{P_{\text{train}}(y)}.
\end{equation}

In order for calibration to be root invariant, I take the mean ratio over chords that share the same quality, and a single calibration factor \(r_q\) is applied to every chord with that quality.

Note that the model's outputs are in logits so the log ratio is added in implementation.

\begin{figure}[H]
    \centering
    \includegraphics[width=0.8\textwidth]{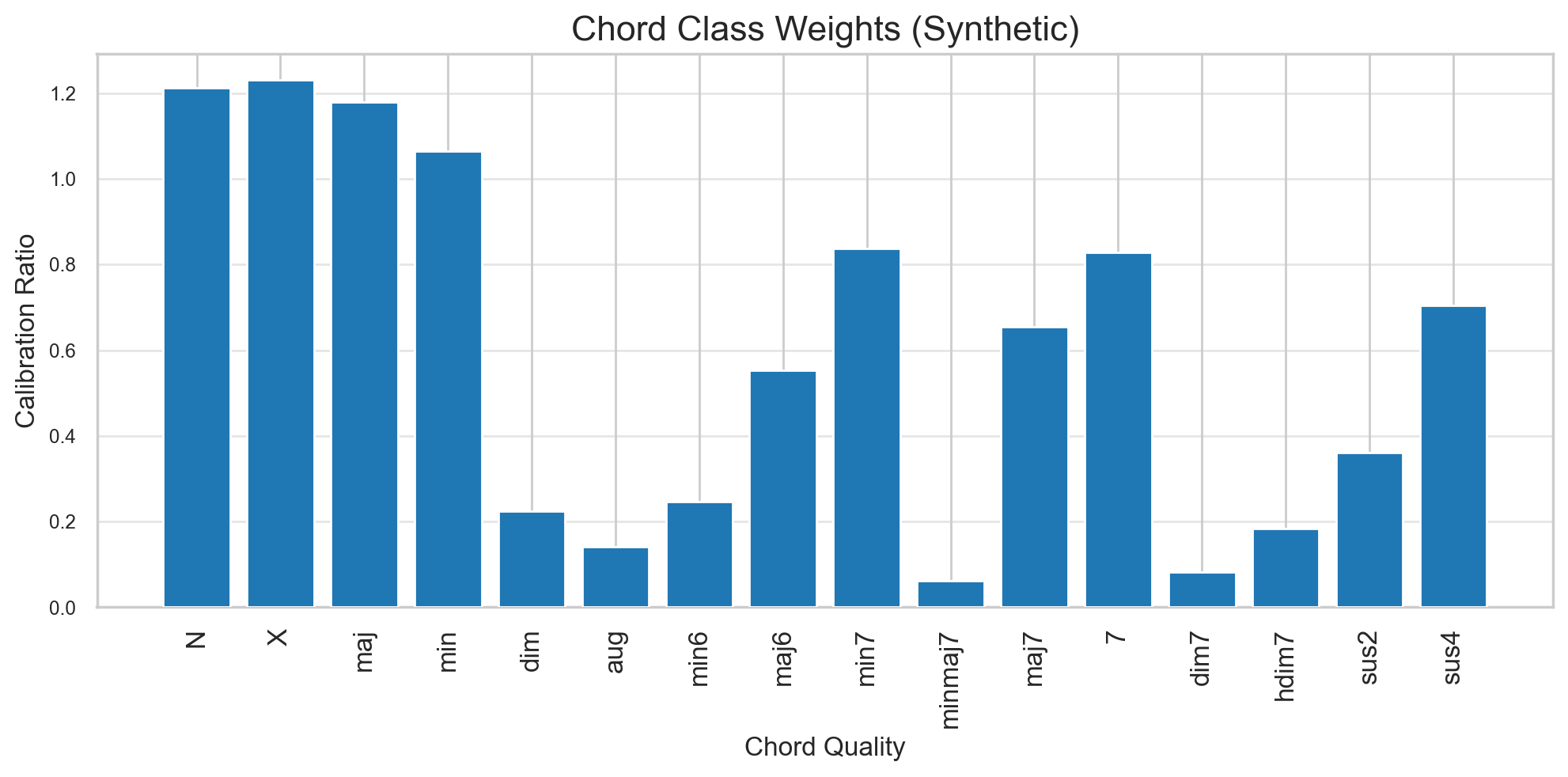}
    \caption{Calibration ratios $r_q$ of the model's outputs to account for distribution shift. The high ratio on rare chord qualities like \texttt{majmin} show that these qualities are much more common in the synthetic dta.}\label{fig:calibration}
\end{figure}

Figure~\ref{fig:calibration} shows the calibration of the model's outputs to account for distribution shift.

\section{Difference in Confusion Matrices with Synthetic Data}\label{app:cm_synthetic_data}

\begin{figure}[H]
    \centering
    \includegraphics[width=0.8\textwidth]{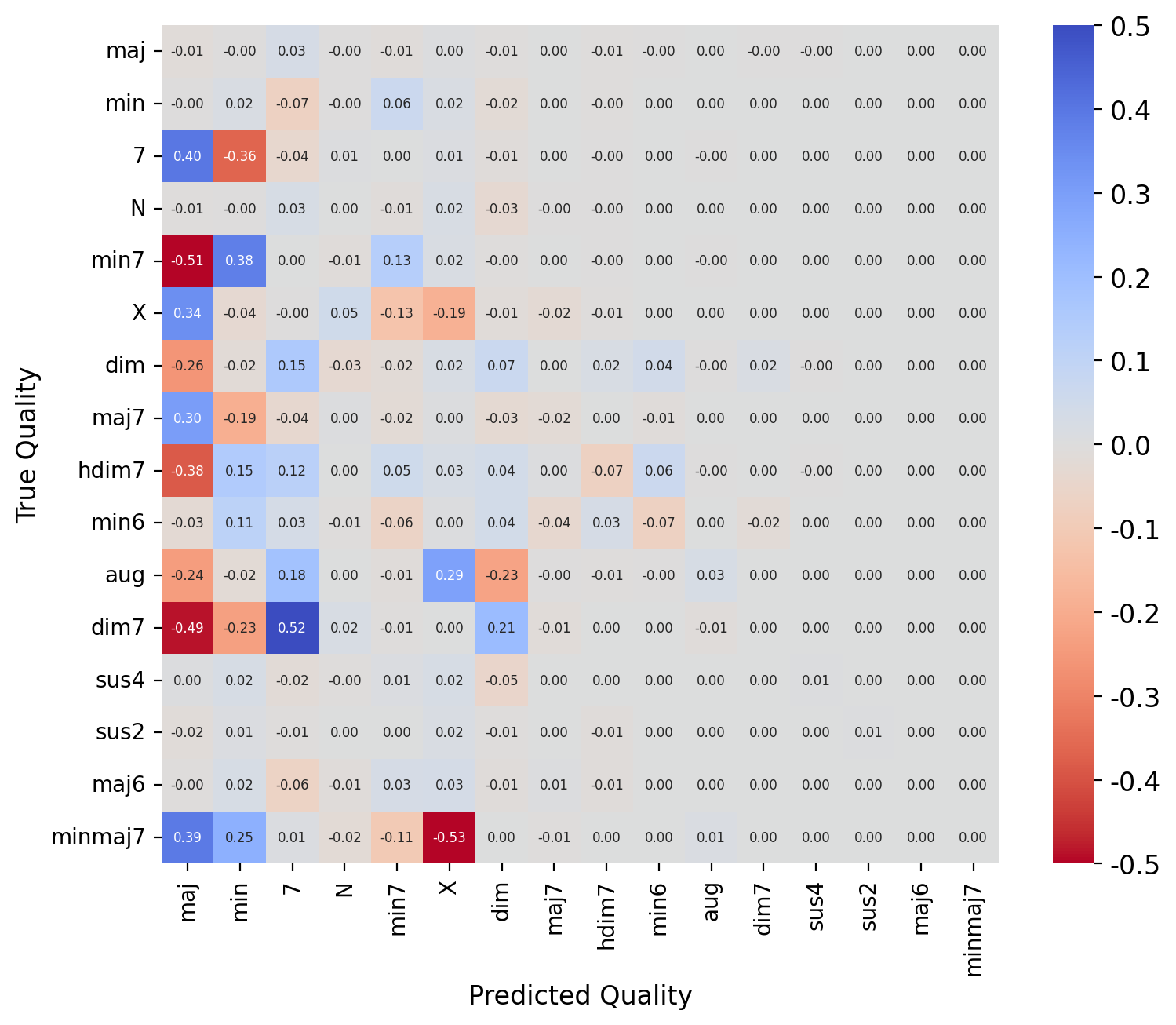}
    \caption{Element-wise difference in normalised confusion matrix of the model trained on synthetic data and \emph{pop} data versus just \emph{pop} data with a weighted loss function. There are some notable differences. Training on synthetic data strongly discourages the model from predicting \texttt{X} chords, which are not present in the synthetic data. Recall on \texttt{min7} qualities increases by $13\%$ and by $7\%$ on \texttt{dim} qualities. However, recall \texttt{hdim7} and \texttt{min6} worsens. In general, the model predicts \texttt{maj} less often for rarer qualities. Another interesting observation is that the synthetic data corrects many of the predictions on \texttt{7} qualities from erroneous predictions of \texttt{min} to erroneous predictions of \texttt{maj}. Something similar happens with \texttt{min7} qualities. It is hard to say which model is better. Indeed, their overall accuracies are the same.}\label{fig:cm_synthetic_data}
\end{figure}

\section{Maximum Lag Cross Correlation of Chord Transitions with Beats}\label{app:maximum_lag_cross_correlation_beats}

\begin{figure}[H]
    \centering
    \includegraphics[width=0.8\textwidth]{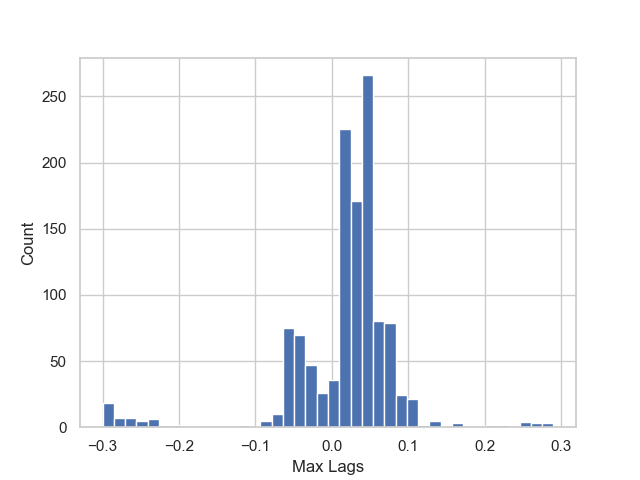}
    \caption{Maximum lag cross correlation of chord transitions with beats within a window of $0.3$ seconds. Almost all maximum lags occur between $-0.1$ and $0.1$ seconds.}\label{fig:maximum_lag_cross_correlation}
\end{figure}